\documentclass[useAMS]{mn2e}
\usepackage{graphicx}
\usepackage{epsfig}
\usepackage{amssymb}
\usepackage{lscape}
\usepackage{ulem}
\usepackage{amsmath}

\def\kms{km~s$^{-1}$}

\def\c2s{C\,{\sc ii}$^{\star}$}

\def\fgas{$f_{gas}$}

\title[The HI properties of post-merger galaxies] {The neutral gas content 
of post-merger galaxies.}

\author[Ellison et al.] {Sara L. Ellison$^1$,
Derek Fertig$^2$,
Jessica L. Rosenberg$^2$, 
Preethi Nair$^3$, 
Luc Simard$^4$,
\newauthor 
Paul Torrey$^5$,
David R. Patton$^6$\\
$^1$ Department of Physics \& Astronomy, University
of Victoria, Finnerty Road, Victoria, British Columbia, V8P 1A1,
Canada.\\
$^2$ School of Physics, Astronomy, and Computational Science,
George Mason University,  MS 3F3, 4400 University Drive, \\
Fairfax, VA 22030, USA.\\
$^3$ Dept. of Physics \& Astronomy, University of Alabama, 514 University Blvd, Tuscaloosa, Alabama, 35487, USA\\
$^4$ National Research Council of Canada, 5071 West Saanich Road, Victoria, BC V9E 2E7, Canada\\
$^5$ Harvard Smithsonian Center for Astrophysics, 60 Garden Street, Cambridge, MA 02138, USA\\
$^6$ Department of Physics \& Astronomy, Trent University,
1600 West Bank Drive, Peterborough, Ontario, K9J 7B8, Canada.
}
\begin{document}

\maketitle

\begin{abstract}

Measurements of the neutral hydrogen gas content of a sample of 93
post-merger galaxies are presented, from a combination of matches
to the ALFALFA.40 data release and new Arecibo observations.
By imposing completeness thresholds identical to that of the ALFALFA
survey, and by compiling a mass-, redshift- and environment-matched control sample
from the public ALFALFA.40 data release, we calculate gas fraction offsets ($\Delta
f_{gas}$) for the post-mergers, relative to the control sample.  We find that 
the post-mergers have HI gas fractions that are consistent with undisturbed galaxies.
However, due to the relative gas richness of the ALFALFA.40 sample,
from which we draw our control sample, our measurements of gas fraction enhancements
are likely to be conservative lower limits.  Combined with comparable gas fraction 
measurements by Fertig et al. in a sample of galaxy pairs, who also determine
gas fraction offsets consistent with zero, we conclude that there is 
no evidence for significant neutral gas consumption
throughout the merger sequence.  From a suite of 75 binary merger simulations
we confirm that star formation is expected to decrease
the post-merger gas fraction by only 0.06 dex, even several Gyr after the merger.
Moreover, in addition to the lack of evidence for gas consumption from gas fraction
offsets, 
the observed HI detection fraction in the complete sample of post-mergers
is twice as high as the controls, which suggests that the post-merger gas fractions 
may actually be enhanced.  We demonstrate that a gas fraction enhancement in post-mergers, 
relative to a stellar mass-matched control sample, would indeed be the
natural result of merging randomly drawn pairs from
a parent population which exhibits a declining gas fraction with increasing
stellar mass.
\end{abstract}


\section{Introduction}

Galaxies require gas for star formation. There exist several
ways in which this gas can be acquired, such as feeding through 
intergalactic filaments, accretion of satellites and major mergers
(e.g. Keres et al. 2005; Wang et al. 2011;
L'Huillier, Combes \& Semelin 2012; and see Sancisi et al.
2008 and Sanchez Almeida et al.  2014 for 
reviews). There is compelling evidence that the gas reservoir
must be continually replenished if star formation is to continue
(see Putman, Peek \& Joung 2012 for a review and discussion of
various refuelling mechanisms).  For example, whilst the
total stellar mass of galaxies on the red sequence increases
significantly towards lower redshifts, both the mass in
galaxies in the `blue cloud' (Bell et al. 2004, 2007; Borch et al.
2006; Faber et al. 2007) 
and the mass density of
HI in high column density QSO absorbers (e.g. Guimaraes et al. 2009;
Noterdaeme et al. 2012; Zafar et al. 2013) exhibit no strong evolution
with redshift.  This lack of evolution contrasts with the star formation
rate (SFR) density, which evolves strongly over the epochs probed by 
these observations
(e.g. Lilly et al. 1996; Madau et al. 1998; Hopkins \& Beacom 2006).
Refuelling of the galactic gas reservoir seems inevitable and
is both widely predicted by simulations and supported by observations 
(e.g. Keres et al 2005; Oosterloo et al. 2007a; Sancisi 
et al. 2008; Brooks et al. 2009).

Galaxy mergers provide a natural way to deliver a fresh supply of gas
to galaxies (although perhaps not sufficient to maintain current
rates of star formation, e.g. Di Teodoro \& Fraternali 2014).  
This delivery mechanism usually triggers
new episodes of star formation in both the pre-coalescence
phase  (Barton, Geller \& Kenyon 2000; Nikolic, Cullen \& Alexander 2004;
Alonso et al. 2006; Woods, Geller \& Barton 2006; 
Ellison et al. 2008a, 2013b; Li et al 2008; Scudder
et al. 2012; Patton et al. 2013) and the final merger remnant  
(Wang et al. 2006; Karteltepe et al. 2010; Daddi et al. 2010a;
Genzel et al. 2010; Carpineti et al. 2012; Kaviraj et al. 2012;
Shabala et al. 2012; Ellison et al. 2013a; Kaviraj 2014).  
Such bursts of star formation may plausibly lead to a more rapid
depletion of the gas reservoir.  
However, observational results on this topic have
yielded mixed results.  Braine \& Combes (1993) found the HI content of
disturbed galaxies to be similar to that of unperturbed galaxies,
and concluded that the HI content was unaffected by intereactions.
In contrast, observations of HI in post-starburst
galaxies, which are implicated as merger-driven (Zabludoff
et al. 1996; Blake et al. 2004), find HI gas fractions that
are low relative to late-types (presumably the merger progenitors)
of the same stellar mass (Zwaan et al. 2013). However, compared
to early types, the post-starbursts are relatively HI rich
(Zwaan et al. 2013). 

In addition to supplying fresh gas, the merger may
also re-distribute the gas, potentially making it unavailable
for sustained star formation.  For example, maps of HI in nearby galaxy pairs
and post-mergers reveal long tidal gas tails and bridges that can extend
up to a few hundred kpc (Hibbard \& Yun 1999; Wang et al. 2001;
Koribalski \& Dickey 2004).  Gas-rich ellipticals, which may be
merger remnants, sometimes exhibit large HI masses distributed
in extended structures, but with surface densities that are
insufficient to form new stars (Oosterloo et al. 2007b). Indeed,
it has been suggested that the long lifetimes of such extended
HI structures, which may have no (or very low surface brightness) 
optical counterparts (Hibbard \& Yun 1999; Buyle et al. 2008)
may be an effective way of identifying mergers long after optical 
disturbances have faded (Hibbard \& van Gorkom 1996; Holwerda et al. 2011;
Lelli, Verheijen \& Fraternali 2014).

Despite these numerous individual studies, larger samples,
coupled with a comparison to isolated (i.e. non-merger)
galaxies of similiar properties,
are required to form a statistical picture of the evolution of
gas content during the merger sequence.
In Fertig et al. (2015) we investigated the evolution
of HI gas fractions in the pre-merger phase of a sample of
close galaxy pairs.  Despite well documented SFR enhancements
in this same sample (Scudder et al. 2012; Patton et al. 2013),
Fertig et al. (2015) do not detect any decline in the HI gas fraction
of close pairs.  Indeed, in an earlier study based on a small sample 
of confusion-free
HI measurements, Huchtmeier et al. (2008) suggested that mergers
may be relatively HI-rich.  Casasola et al. (2004)
present a similar study, but with over 1000 galaxies
drawn from all interaction stages, from well separated pairs to
coalesced remnants.  HI gas fractions are again
found to be enhanced by a factor of
$\sim$ 10 in early type systems, but much more modestly over-abundant
in late type galaxies.

There are a few possible reasons why previous studies have not detected
depletion in merger gas fractions.   First, in order to obtain
a complete view of gas depletion, it is important to
consider both atomic and molecular gas, the latter of 
which may be expected to be more closely linked to star formation.
One of the first studies of the combined atomic and molecular
content of galaxy mergers was performed by Braine \& Combes (1993), in
which it was found that the molecular gas fraction is enhanced
in interacting galaxies.
Bettoni et al. (2001), from a study of the CO and HI content of polar
ring galaxies (which have likely formed through a past merger), also
found  gas fractions that are enhanced by
an order of magnitude.  Enhanced molecular gas fractions soon after
a merger are predicted by simulations (Bekki 2014), although eventually
the gas is consumed by the triggered star formation.   
Stark et al. (2013) measured elevated molecular fractions in galaxies
selected to have anomolously blue centres, which are frequently the
signature of a recent merger.  A similar result is found by
Goncalves et al. (2014) who measure elevated CO fractions in their local
sample of Lyman Break Analogs, many of which show merger properties.
Ultra-luminous infra-red galaxies (ULIRGs), which are often merger
driven, also manifest very high CO gas fractions
(Daddi et al. 2010b). Most recently, Ueda et al. (2014) have used a sample of
37 post-mergers to show that molecular gas disks are a common feature 
of merger remnants, indicating that these reservoirs have not
yet been exhausted by merger driven star formation.

A second potential reason for the lack of HI depletion in mergers 
is the relatively small consumption expected in the pre-merger
(pairs) phase.  Fertig et al. (2015) used a suite of binary merger
simulations to show that, even at the closest separations, the
HI gas mass is expected to be depleted by only $\sim$ 0.04 dex.  
This is a relatively 
small effect that can be challenging to detect in small samples.
Signatures of gas depletion may be more obvious in late-stage mergers, where
simulations predict that SFRs will peak (e.g. Mihos \& Hernquist 1996; 
Di Matteo et al 2007, 2008; Montuori et a. 2010; Torrey et al. 2012). 
In Ellison et al. (2013a) we confirmed 
the prediction of high SFRs by combining a sample of galaxy pairs
(with separations up to 80 kpc) and post-mergers.  There is
a steady increase in the SFR enhancements with decreasing separation,
peaking in the post-mergers.  We may therefore expect that the most
rapid consumption of gas occurs after coalescence, depending on the
rate of star formation and its duration. Indeed, simulations of 
galactic gas reservoirs of both atomic and molecular gas show an increase 
in gas fraction before a decline at later times 
(Bekki 2014; Rafieferantsoa et al. 2014). The initial increase
in HI gas fraction may be linked to the condensation of hot halo
gas into the disk (Moster et al. 2011; Tonnesen \& Cen 2012).
High molecular gas fractions may follow from the increased central
surface densities that precede starbursts in mergers.
Indeed, several previous observational studies have measured
increased gas fractions in either atomic or molecular gas (Braine
\& Combes 1993; Bettoni et al. 2001; Huchtmeier et al. 2008;
Stark et al. 2013; Goncalves et al. 2014).

In this paper, we present a systematic study of HI masses in
a moderately large sample of 93 recently merged galaxies.  
The sample is constructed from both new Arecibo observations,
combined with publicly available HI masses measured as part of
the Arecibo Legacy Fast ALFA (ALFALFA) project (Giovanelli 2005a,b;
Saintonge 2007).  Through
a careful matching of the post-mergers with a sample of isolated (non-merger)
galaxies, it is possible to quantify the enhancement or deficit in 
gas fraction associated with the merger process.  Finally, we interpret
our results with the aid of hydrodynamical merger simulations.

We adopt a cosmology of  $\Omega_{\Lambda} = 0.7$, $\Omega_M = 0.3$,
$H_0 = 70$ km/s/Mpc.

\section{Samples and observations}\label{obs_sec}

\subsection{The Ellison et al. (2013a) Galaxy Zoo post-merger sample}

Ellison et al. (2013a) identified and studied the optical properties of 
97 post-merger galaxies,
selected from the SDSS, based on the visual classifications of the Galaxy
Zoo project (Darg et al. 2010).  In brief, the Galaxy Zoo visual
identifications are performed on some 900,000 galaxies in the
SDSS DR6, as described in Lintott et al. (2008).  For the compilation
of their merger samples, Darg et al. (2010) additionally required 0.0005
$< z <$ 0.1 and $m_r < 17.77$.  Further visual classification
was performed by Ellison et al. (2013a) with the additional requirements
that galaxies have an SDSS specclass=2 and that a Mendel et al. (2014)
stellar mass must be available.  The final sample is characterized
by single late stage mergers that are highly disturbed (i.e. these are
not galaxy pairs), with distinctive shells and tidal features and
is therefore likely to be dominated by the remnants of gas-rich major
mergers.  Although it is not possible to know the exact mass ratio
of the mergers that formed these remnants, simulations indicate that
the mass ratios were likely closer than 1:4 (e.g. Lotz et al. 2010b;
Ji, Peirani \& Yi 2014).

\subsubsection{Matches with ALFALFA.40}

\begin{table*}
\begin{center}
\caption{HI masses from the ALFALFA.40 sample matched to post-mergers in the Ellison et al. (2013a) Galaxy Zoo sample.}
\begin{tabular}{lrrrrrrrrrr}
\hline
         SDSS   objid   &  AGC \# &  R.A.   &   dec.   &  redshift   &   log M$_{\star}$   &  W$_{50}$ &    S$_{21}$ &  RMS  &  S/N &    log M(HI)\\	
                        &         &          &         &             & (M$_{\odot}$)  & (\kms)      &  (Jy \kms)   &  (mJy) &       &   (M$_{\odot}$)\\
 \hline	
588010880366477449 & 212593 &  172.62860 &    5.8920 &  0.03493  &  10.26 & 279$\pm$27 & 1.86$\pm$0.12 & 2.51 & 9.90 & 10.02$\pm$0.03 \\  
587736479206342888 & 714739 &  228.22300 &   10.1099 &  0.03740  &   8.90 & 133$\pm$9 & 0.83$\pm$0.13 & 2.10 & 8.60 &  9.70$\pm$0.07 \\  
587739132428157191 & 727297 &  237.06779 &   25.5270 &  0.04173  &  10.06 & 271$\pm$40 & 1.19$\pm$0.10 & 2.34 & 6.90 &  9.96$\pm$0.04 \\  
587736543096799321 & 253937 &  226.32449 &    8.1540 &  0.03912  &  10.01 & 151$\pm$31 & 0.96$\pm$0.07 & 2.24 & 7.80 &  9.82$\pm$0.03 \\  
587724233182412906 & 113181 &   28.62170 &   14.1960 &  0.05065  &   9.64 & 98$\pm$36 & 1.35$\pm$0.06 & 2.33 & 12.90 & 10.16$\pm$0.02 \\  
587739382067822837 & 727193 &  234.00630 &   25.5510 &  0.03616  &  10.44 & 319$\pm$19 & 1.33$\pm$0.10 & 2.35 & 7.10 &  9.88$\pm$0.03 \\  
587736808845082795 & 245068 &  223.05690 &   12.0609 &  0.05250  &  10.48 & 291$\pm$7 & 2.01$\pm$0.09 & 2.16 & 12.20 & 10.39$\pm$0.02 \\  
588010880378404942 & 8372 &  199.99110 &    5.8080 &  0.02134  &  10.51 & 346$\pm$10 & 2.89$\pm$0.10 & 2.23 & 15.50 &  9.79$\pm$0.01 \\  
\hline
\end{tabular}
\label{alf_tab}
\end{center}
\medskip
The AGC number refers to the ALFALFA catalog number.  W$_{50}$ is the FWHM
of the 21 cm emission line, S$_{21}$ is the integrated 21 cm
line flux.  Derivations of the RMS and S/N are given in Haynes et al. (2011). Stellar masses are taken
from the single Sersic fits of Mendel et al. (2014). For objID=587736479206342888
the spectral parameters (including W$_{50}$, S$_{21}$, the RMS and final derived HI
mass) are determined from our own re-analysis of the ALFALFA data.
\end{table*}

The Ellison et al. (2013a) post-merger sample was cross-correlated
with the ALFALFA survey (Giovanelli et al. 2005a,b).
 ALFALFA is a large survey conducted with the Arecibo telescope covering
$\sim$ 7000 square degrees of sky, with a frequency coverage between 1335 and 
1435 MHz, i.e. with coverage of the 21 cm line
extending to $z \sim 0.06$.  The survey is conducted in
a drift scan mode, such that sensitivity is naturally dependent on
the galaxy distance.  Giovanelli et al. (2005a) estimate the HI mass
limit to be $\sim 10^6$ M$_{\odot}$ at a distance of 6.5 Mpc and
$10^7$ M$_{\odot}$ at 20 Mpc.  The 40\% data release 
(ALFALFA.40, Haynes et al. 2011) covers $\sim$ 2800 square degrees and
contains $\sim$ 15,000 extragalactic sources.  For the cross-matching, 
we searched for ALFALFA.40 detections within a 1.5 arcmin radius and 
a difference in recessional velocity of less than 500 km/s for each 
post-merger. 
In practice, most matches were successfully made with offsets $<$
30 arcseconds.  This procedure
yielded 8 matches, see Table \ref{alf_tab}.  Of these 8, 7 are flagged as
high quality detections in ALFALFA.40 (identified by the survey as  code 1). 
In addition, there is one SDSS post-merger (objID=587736808845082795)
that is associated with an ALFALFA.40 source that is flagged as a code
2 galaxy\footnote{The ALFALFA catalog allocates a code 2 flag for galaxies 
with detections whose S/N$<$6.5. or with otherwise
noisy spectra.  In the case of objID=587736808845082795, the data
cube is significantly affected by RFI, hence the classification as
`code 2', despite the relatively high formal S/N.}.  

For one object,
objID=587736479206342888, the ALFALFA.40 catalog reports a suspiciously
low velocity width of 27 \kms.  Upon inspection of the spectrum, the data
were found to be very noisy with a possible second component in the emission.
The original spectrum only measured one narrow peak resulting in a under 
estimate of the velocity width. We applied further smoothing to the 
spectrum and re-measured the flux adding the second peak.  From this
spectrum we then re-computed the parameters for this galaxy with our own
data reduction tools, and report the final adopted value in Table \ref{alf_tab}.

\subsubsection{Targeted Arecibo observations}

\begin{figure*}
\centerline{\rotatebox{0}{\resizebox{16cm}{!}
{\includegraphics{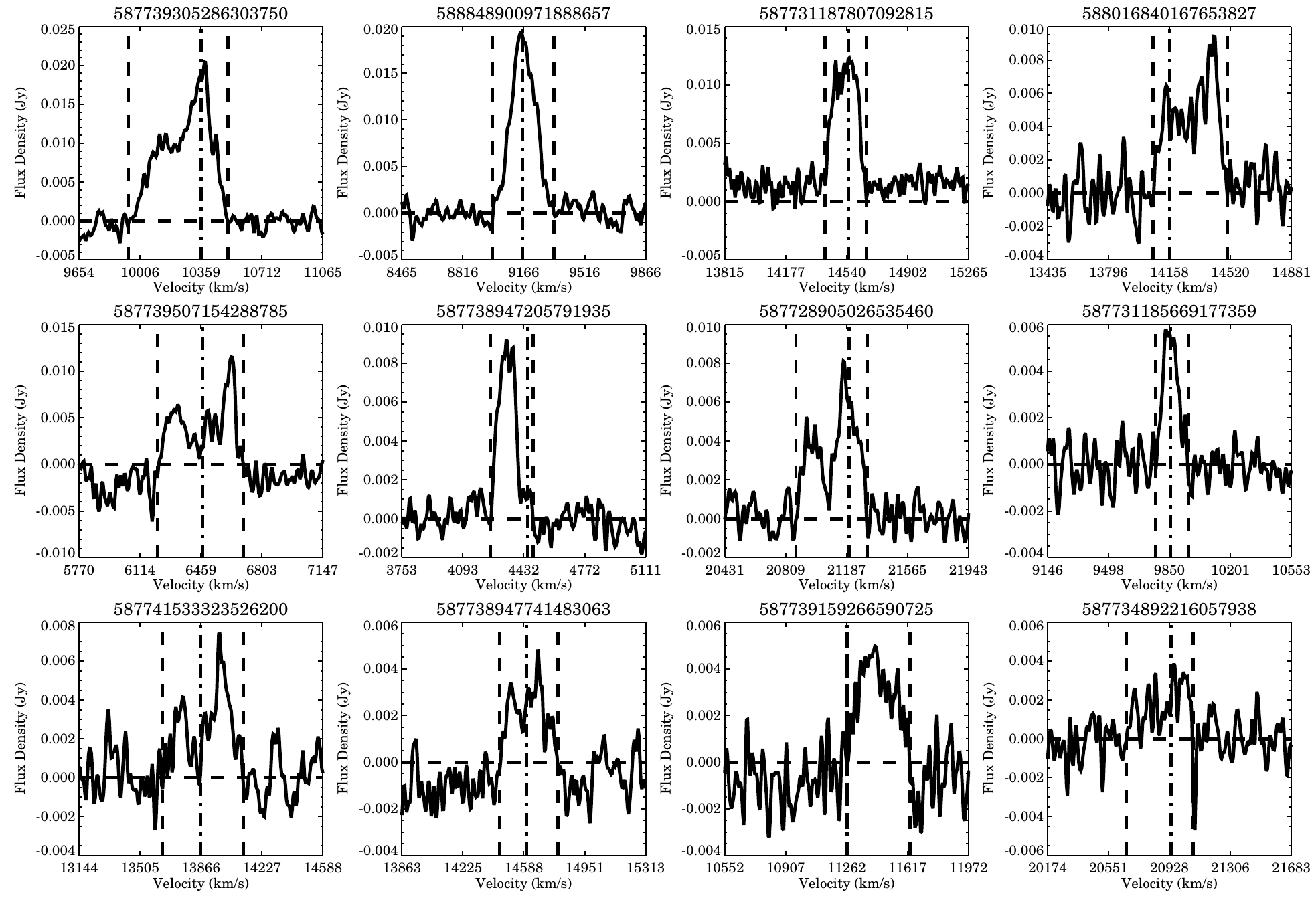}}}}
\caption{\label{spectra} Spectra of the 21 cm emission lines for
the 12 new detections in our post-merger sample (see Table \ref{PI_tab}). 
The vertical dashed lines indicate the velocity range over which the
flux is summed and the vertical dot-dash line shows the redshift derived from
the optical spectrum.  The SDSS objID is given above each panel.}
\end{figure*}

\begin{table*}
\begin{center}
\caption{Target list and parameters for Aug. and Dec 2013 Arecibo observations.}
\begin{tabular}{lrrrrrrrrrr}
\hline
         SDSS   objid   &     R.A.   &   dec.   &  redshift   &    log M$_{\star}$ &  Time  &   W$_{50}$ & S$_{21}$ &  RMS  &  S/N &     log M(HI)\\	
                        &            &          &              &  (M$_{\odot}$)& (ON+OFF)&  (\kms)  & (Jy \kms)&   (mJy) &      & (M$_{\odot}$)\\  
\hline
587731187807092815 &  343.01901 &    1.2500 &  0.04854  &  10.10 & 1140 & 190$\pm$10 & 2.23$\pm$0.29 & 2.33 & 15.62 & 10.35$\pm$0.06 \\  
587731185669177359 &    4.87780 &   -0.6020 &  0.03287  &   9.32 & 1800 & 100$\pm$8 & 0.57$\pm$0.07 & 0.89 & 14.14 &  9.42$\pm$0.05 \\  
588016840167653827 &  117.37570 &   18.9580 &  0.04724  &  10.37 & 240 & 351$\pm$16 & 2.08$\pm$0.27 & 1.82 & 13.71 & 10.31$\pm$0.06 \\  
587728905026535460 &  119.24500 &   34.0929 &  0.07073  &  10.05 & 1200 & 342$\pm$12 & 1.62$\pm$0.15 & 1.02 & 19.12 & 10.54$\pm$0.04 \\  
587738947741483063 &  140.58290 &   31.2859 &  0.04871  &   9.72 & 1800 & 227$\pm$15 & 0.74$\pm$0.15 & 1.20 & 9.08 &  9.88$\pm$0.09 \\  
587738947205791935 &  143.69679 &   32.1430 &  0.01486  &   8.52 & 1800 & 122$\pm$6 & 1.07$\pm$0.10 & 0.95 & 22.87 &  8.99$\pm$0.04 \\  
588848900971888657 &  146.79910 &    0.7030 &  0.03055  &   9.44 & 1200 & 170$\pm$9 & 3.20$\pm$0.27 & 2.07 & 26.49 & 10.11$\pm$0.04 \\  
587739159266590725 &  153.54409 &   34.3429 &  0.03758  &   9.99 & 240 & 189$\pm$26 & 1.08$\pm$0.15 & 1.08 & 16.23 &  9.83$\pm$0.06 \\  
587739305286303750 &  172.32220 &   35.5769 &  0.03457  &  10.96 & 360 & 369$\pm$15 & 5.29$\pm$0.42 & 2.66 & 23.15 & 10.43$\pm$0.03 \\  
587741533323526200 &  173.78129 &   29.8910 &  0.04624  &  11.07 & 240 & 333$\pm$10 & 1.25$\pm$0.16 & 1.25 & 12.24 & 10.06$\pm$0.06 \\  
587739507154288785 &  185.06550 &   33.6609 &  0.02157  &  10.46 & 240 & 414$\pm$6 & 2.06$\pm$0.26 & 1.98 & 11.18 &  9.62$\pm$0.05 \\  
587734892216057938 &  179.54049 &    9.5360 &  0.06985  &  10.69 & 240 & 147$\pm$25 & 0.71$\pm$0.12 & 2.31 & 5.66 & 10.17$\pm$0.07 \\  
588015510343712908 &   20.46209 &    0.8470 &  0.09898  &  11.15 & 240 & 438  & $<$0.91  & 1.43 & 6.5 & $<$10.59  \\  
587730775492001995 &  342.48558 &   14.8750 &  0.08993  &  10.69 & 540 & 320  & $<$0.42  & 0.81 & 6.5 & $<$10.17  \\  
587731185651810753 &  325.23431 &   -0.4780 &  0.06816  &   9.74 & 1800 & 185  & $<$0.51  & 1.28 & 6.5 & $<$10.01  \\  
587738372206690585 &  120.58889 &   19.9939 &  0.06771  &  10.58 & 240 & 359  & $<$0.91  & 1.65 & 6.5 & $<$10.26  \\  
587732470387703859 &  129.57679 &   33.5779 &  0.06212  &  11.18 & 240 & 470  & $<$1.45  & 2.13 & 6.5 & $<$10.38  \\  
587732484342415393 &  130.93750 &   35.8279 &  0.05393  &  11.04 & 240 & 373  & $<$0.99  & 1.75 & 6.5 & $<$10.09  \\  
587744729564512606 &  133.78070 &   15.2690 &  0.07681  &  10.99 & 240 & 523  & $<$1.07  & 1.41 & 6.5 & $<$10.44  \\  
587738409785557168 &  143.44760 &   10.8120 &  0.08558  &  11.33 & 240 & 369  & $<$0.77  & 1.38 & 6.5 & $<$10.39  \\  
587735349633351728 &  148.30130 &   13.1009 &  0.07709  &  11.34 & 240 & 374  & $<$0.83  & 1.47 & 6.5 & $<$10.33  \\  
587738947747053602 &  154.64019 &   36.2239 &  0.05412  &  10.95 & 240 & 439  & $<$1.49  & 2.33 & 6.5 & $<$10.27  \\  
587726032776265859 &  159.13279 &    2.3620 &  0.05026  &  10.30 & 240 & 279  & $<$1.15  & 2.36 & 6.5 & $<$10.10  \\  
587734948595236905 &  160.26559 &   11.0959 &  0.05297  &  10.82 & 240 & 434  & $<$0.83  & 1.32 & 6.5 & $<$10.00  \\  
587732702329045015 &  160.29069 &    6.8200 &  0.03332  &  10.57 & 240 & 344  & $<$0.88  & 1.63 & 6.5 & $<$ 9.63  \\  
588848900441112762 &  160.79060 &    0.3140 &  0.09788  &  11.01 & 240 & 590  & $<$1.77  & 2.07 & 6.5 & $<$10.87  \\  
588017702386270301 &  162.93029 &   10.7570 &  0.09734  &  11.52 & 240 & 448  & $<$0.83  & 1.28 & 6.5 & $<$10.53  \\  
588017703997276265 &  163.71000 &   12.0900 &  0.08045  &  10.55 & 480 & 244  & $<$0.53  & 1.17 & 6.5 & $<$10.17  \\  
588848900981260449 &  168.23460 &    0.7260 &  0.09687  &  11.24 & 240 & 348  & $<$0.94  & 1.73 & 6.5 & $<$10.58  \\  
587741533859414124 &  171.14239 &   30.0960 &  0.05492  &  10.74 & 240 & 257  & $<$0.71  & 1.53 & 6.5 & $<$ 9.97  \\  
587742013816897595 &  175.26300 &   22.0480 &  0.06290  &   9.91 & 960 & 297  & $<$0.65  & 1.30 & 6.5 & $<$10.05  \\  
587742062680015040 &  176.18339 &   23.1620 &  0.04839  &  10.58 & 240 & 415  & $<$0.95  & 1.57 & 6.5 & $<$ 9.98  \\  
587742062680997918 &  178.51669 &   23.3400 &  0.05093  &  10.16 & 240 & 426  & $<$0.80  & 1.29 & 6.5 & $<$ 9.95  \\  
587726031175221368 &  180.99819 &    1.4110 &  0.02230  &  10.18 & 240 & 205  & $<$0.58  & 1.39 & 6.5 & $<$ 9.10  \\  
588017703469777068 &  185.68820 &   12.2939 &  0.09209  &  11.54 & 240 & 568  & $<$1.62  & 1.96 & 6.5 & $<$10.77  \\  
588017730842853516 &  201.19819 &    8.2950 &  0.05179  &  10.94 & 240 & 293  & $<$0.85  & 1.70 & 6.5 & $<$ 9.99  \\  
587741603112157297 &  201.27389 &   27.5450 &  0.07209  &  10.85 & 240 & 405  & $<$0.94  & 1.59 & 6.5 & $<$10.32  \\  

\hline
\end{tabular}
\label{PI_tab}
\end{center}
\medskip
The AGC number refers to the ALFALFA catalog number.  W$_{50}$ is the FWHM
of the 21 cm emission line for detections and calculated from the 
$i$-band Tully Fisher relation for non-detections (Giovanelli et al. 1997). 
S$_{21}$ is the integrated 21 cm
line flux.  Derivations of the RMS and S/N are given in Haynes et al. 
(2011).  The S/N is set to
6.5 for all non-detections. Stellar masses are taken from
the single Sersic fits of Mendel et al. (2014). 
\end{table*}

A further 37 post-mergers were observed during two runs (PI Fertig)
with the Arecibo telescope on August 19 and December 28, 29 2013,  
see Table \ref{PI_tab}.  The observations used the L-band 
wide receiver with  the Interim Correlator and 9-level sampling with 
2 polarizations per board. As for the ALFALFA survey, we used the
WAPP spectrometers with a total bandwidth of 100MHz centered on 1385 Mhz.
The observations were designed to reach a 5$\sigma$ H I mass limit of 10\% 
of the stellar mass.  The minimum exposure time for the ON+OFF was set 
to be 4 minutes. For four of the sources 
the observing time was capped at 30 minutes, yielding a 5$\sigma$ HI mass limit 
of 50\% of the stellar mass or better.  
 For the data reduction, the spectra were Hanning smoothed and then reduced 
using standard ON-OFF reduction. Five sources were observed on 19 August -- two 
were detections and three were non-detections. An additional 32 sources were 
observed on 28 and 29 December resulting in 10 more detections.  Reduced
spectra of the 12 new detections are shown in Fig. \ref{spectra}.  The
final resolution ranges between 10 and 14 \kms, depending on smoothing.

Table \ref{PI_tab} lists various target and observational parameters, including
exposure times, fluxes and the final derived HI masses for our Arecibo 
observations.
HI masses (in units of solar masses) are calculated using the following 
formula:

\begin{equation}\label{m_eqn}
M(HI)=2.356 \times 10^5 \times D^2 S_{21}
\end{equation}

where D is the distance to the source in Mpc ($cz/H_0$, consistent with
ALFALFA) and $S_{21}$ is the 21 cm line flux in Jy \kms.
All detections have S/N $>$ 5, with 11/37 of the 
spectra having S/N$>$6.5 (the requirement for a
code 1 classification in the ALFALFA.40 catalog).   
The limiting flux for non-detections is calculated from

\begin{equation}\label{s_eqn}
S_{21} = S/N \times \frac{\rm RMS}{w^{0.5}_{sm}}\frac{W_{50}}{1000}
\end{equation}

where the RMS is measured from the spectrum and $w_{sm}$ is a smoothing width. 
This latter parameter is either $W_{50}$/20 
for $W_{50} < 400$ km/s or 400/20=20 for $W_{50} > 400$km/s. For 
non-detections, the full width at half maximum (FWHM)
line width $W_{50}$ is calculated from the $i$-band Tully Fisher relation
(Giovanelli et al. 1997).  
The 21 cm flux limits for the non-detections were then calculated from
the inferred $W_{50}$ values inserted into 
equation \ref{s_eqn} for a limiting S/N (significance) 
of 6.5 in order to be consistent with values quoted
from ALFALFA.  The non-detections in Table \ref{alf_tab} therefore have
the S/N fixed to a value of 6.5. 

Combined with the ALFALFA.40 matches, we therefore have
a total of 45 HI mass measurements for the Ellison et al. (2013a)
Galaxy Zoo post-merger sample, of which 20 are detections and 25 are
upper limits.

\subsection{The Nair \& Abraham (2010) supplementary sample.}\label{supp_sec}

\begin{table*}
\begin{center}
\caption{HI masses from the ALFALFA.40 sample matched to post-mergers selected from Nair \& Abraham (2010)}
\begin{tabular}{lrrrrrrrrrr}
\hline
         SDSS   objid   &  AGC \# &  R.A.   &   dec.   &  redshift   &  log  M$_{\star}$   &  W$_{50}$ &    S$_{21}$ &  RMS  &  S/N &    log M(HI)\\	
                        &         &          &         &             & (M$_{\odot}$)  & (\kms)      &  (Jy \kms)   &  (mJy) &       &   (M$_{\odot}$)\\          
\hline
587724232100282512 & 100318 &  8.77704 &  14.3548 &  0.03801  &  10.02 & 299$\pm$22 & 2.23$\pm$0.10 & 2.37 & 12.1 & 10.12$\pm$0.02 \\  
587724232641937419 & 110216 &  20.011 &  14.3618 &  0.03104  &  10.12 & 83$\pm$13 & 1.13$\pm$0.06 & 1.94 & 14.2 &  9.66$\pm$0.02 \\  
587724233720332316 & ... &  30.9965 &  14.3104 &  0.04269  &  10.92 & 384  & $<$1.28  & 2.25 & 6.5 & $<$10.00  \\  
587726101750546619 & 231606 &  210.551 &  4.58573 &  0.04031  &  11.02 & 421$\pm$27 & 1.68$\pm$0.12 & 2.44 & 7.3 & 10.09$\pm$0.03 \\  
587727221949530261 & ... &  357.996 &  14.7511 &  0.04668  &  10.95 & 408  & $<$1.33  & 2.25 & 6.5 & $<$10.10  \\  
587727223020454084 & ... &  351.309 &  15.2448 &  0.04342  &  10.40 & 276  & $<$1.09  & 2.25 & 6.5 & $<$ 9.95  \\  
587728880873701381 & ... &  146.112 &  4.49914 &  0.04671  &  10.58 & 408  & $<$1.33  & 2.25 & 6.5 & $<$10.10  \\  
587728881413390384 & 201303 &  152.655 &  5.15067 &  0.01367  &   9.10 & 134$\pm$9 & 1.41$\pm$0.07 & 2.32 & 11.7 &  9.11$\pm$0.02 \\  
587728881418502205 & 6049 &  164.38 &  5.69873 &  0.05320  &  11.16 & 290$\pm$3 & 1.98$\pm$0.11 & 2.56 & 10.1 & 10.41$\pm$0.02 \\  
587729158966345769 & 222264 &  194.588 &  4.88565 &  0.03612  &  10.22 & 355$\pm$10 & 1.68$\pm$0.13 & 2.67 & 7.5 & 10.00$\pm$0.03 \\  
587729160042119249 & ... &  199.249 &  5.65679 &  0.03306  &  10.77 & 343  & $<$1.21  & 2.25 & 6.5 & $<$ 9.76  \\  
587730774424223763 & ... &  356.624 &  14.8867 &  0.05751  &  10.77 & 330  & $<$1.19  & 2.25 & 6.5 & $<$10.23  \\  
587732053235794407 & ... &  118.396 &  26.4624 &  0.03645  &  10.18 & 190  & $<$0.90  & 2.25 & 6.5 & $<$ 9.72  \\  
587732577233469472 & ... &  154.101 &  5.6954 &  0.04899  &  10.97 & 325  & $<$1.18  & 2.25 & 6.5 & $<$10.09  \\  
587732577774469219 & 201673 &  163.518 &  6.72539 &  0.02659  &  10.38 & 229$\pm$3 & 1.96$\pm$0.08 & 2.12 & 13.6 &  9.81$\pm$0.02 \\  
587732578837201105 & 191115 &  138.302 &  5.61087 &  0.05846  &  10.99 & 206$\pm$7 & 1.30$\pm$0.10 & 2.71 & 7.4 & 10.30$\pm$0.03 \\  
587732579385933949 & 213056 &  165.496 &  8.10891 &  0.03019  &   9.76 & 219$\pm$12 & 1.78$\pm$0.09 & 2.32 & 11.6 &  9.88$\pm$0.02 \\  
587732702331535424 & ... &  166.047 &  7.06904 &  0.03255  &  10.71 & 274  & $<$1.08  & 2.25 & 6.5 & $<$ 9.70  \\  
587732702868865044 & ... &  167.004 &  7.44229 &  0.04111  &  10.62 & 374  & $<$1.27  & 2.25 & 6.5 & $<$ 9.97  \\  
587732703398527109 & ... &  150.393 &  6.93203 &  0.04892  &  11.31 & 428  & $<$1.40  & 2.25 & 6.5 & $<$10.16  \\  
587732771053830198 & ... &  178.604 &  9.60905 &  0.03517  &  10.55 & 329  & $<$1.19  & 2.25 & 6.5 & $<$ 9.80  \\  
587732771053830214 & 210905 &  178.637 &  9.54455 &  0.03576  &   9.70 & 180$\pm$11 & 1.06$\pm$0.09 & 2.53 & 7.0 &  9.80$\pm$0.03 \\  
587732772126785620 & 210781 &  176.792 &  10.5177 &  0.02137  &   9.57 & 100$\pm$9 & 0.55$\pm$0.06 & 2.21 & 5.5 &  9.08$\pm$0.04 \\  
587732772130652231 & ... &  185.741 &  10.5483 &  0.02594  &  10.54 & 304  & $<$1.14  & 2.25 & 6.5 & $<$ 9.52  \\  
587734622163763258 & 180008 &  120.867 &  25.1027 &  0.02758  &  10.08 & 237$\pm$74 & 0.96$\pm$0.09 & 2.39 & 5.8 &  9.52$\pm$0.04 \\  
587734891683119143 & 220805 &  188.681 &  9.00471 &  0.04306  &  10.67 & 286$\pm$65 & 0.84$\pm$0.10 & 2.44 & 4.5 &  9.85$\pm$0.05 \\  
587734892216713241 & ... &  181.066 &  9.3969 &  0.03525  &  10.64 & 354  & $<$1.23  & 2.25 & 6.5 & $<$ 9.82  \\  
587734893287440431 & 210497 &  174.042 &  10.0556 &  0.02073  &  10.41 & 261$\pm$23 & 2.87$\pm$0.11 & 2.32 & 17.1 &  9.77$\pm$0.02 \\  
587734894357446674 & ... &  165.312 &  10.4811 &  0.04222  &  10.84 & 346  & $<$1.22  & 2.25 & 6.5 & $<$ 9.97  \\  
587734948054696027 & ... &  151.83 &  9.87247 &  0.05735  &  11.17 & 452  & $<$1.48  & 2.25 & 6.5 & $<$10.32  \\  
587735347489341442 & 202805 &  156.547 &  12.5738 &  0.03096  &  10.14 & 292$\pm$14 & 1.77$\pm$0.10 & 2.34 & 9.9 &  9.90$\pm$0.02 \\  
587735348021624886 & ... &  145.944 &  11.4564 &  0.03906  &  10.75 & 270  & $<$1.08  & 2.25 & 6.5 & $<$ 9.85  \\  
587735348025884725 & ... &  155.81 &  12.8733 &  0.03295  &  10.53 & 262  & $<$1.06  & 2.25 & 6.5 & $<$ 9.70  \\  
587735349086912675 & 181028 &  126.597 &  8.42377 &  0.04594  &  11.17 & 205$\pm$11 & 0.62$\pm$0.09 & 2.67 & 3.6 &  9.77$\pm$0.06 \\  
587736478124408901 & ... &  209.255 &  11.6114 &  0.03870  &  10.72 & 332  & $<$1.19  & 2.25 & 6.5 & $<$ 9.89  \\  
587736802935373840 & ... &  196.764 &  13.637 &  0.02727  &  10.23 & 312  & $<$1.16  & 2.25 & 6.5 & $<$ 9.57  \\  
587736813672202319 & ... &  237.478 &  9.82864 &  0.04587  &  10.92 & 404  & $<$1.32  & 2.25 & 6.5 & $<$10.08  \\  
587738409251635215 & 190666 &  150.264 &  11.4614 &  0.03629  &  10.34 & 225$\pm$8 & 3.17$\pm$0.09 & 2.30 & 20.5 & 10.28$\pm$0.01 \\  
587738410860806225 & 190505 &  146.827 &  12.1585 &  0.04751  &  10.75 & 216$\pm$6 & 1.07$\pm$0.08 & 2.25 & 7.3 & 10.04$\pm$0.03 \\  
587738411401412648 & ... &  155.398 &  13.7755 &  0.04554  &  10.91 & 389  & $<$1.29  & 2.25 & 6.5 & $<$10.06  \\  
588010359073931336 & 201281 &  151.836 &  4.07931 &  0.02859  &  10.48 & 358$\pm$10 & 1.28$\pm$0.11 & 2.48 & 6.1 &  9.69$\pm$0.03 \\  
588010360148656136 & 200217 &  154 &  4.95476 &  0.03203  &  10.02 & 325$\pm$6 & 1.72$\pm$0.10 & 2.33 & 9.2 &  9.91$\pm$0.02 \\  
588010879841992790 & 231357 &  201.04 &  5.25972 &  0.02489  &   9.81 & 229$\pm$4 & 2.33$\pm$0.09 & 2.31 & 14.9 &  9.83$\pm$0.02 \\  
588010880371654746 & 220306 &  184.483 &  5.86978 &  0.02347  &   9.72 & 287$\pm$31 & 2.01$\pm$0.11 & 2.57 & 10.3 &  9.71$\pm$0.02 \\  
588017569779744827 & ... &  203.101 &  11.1064 &  0.03146  &  10.06 & 302  & $<$1.14  & 2.25 & 6.5 & $<$ 9.69  \\  
588017702921633970 & ... &  159.469 &  10.7583 &  0.05251  &  10.99 & 432  & $<$1.42  & 2.25 & 6.5 & $<$10.23  \\  
588017702940049506 & 232401 &  202.338 &  11.3365 &  0.04771  &  10.40 & 329$\pm$14 & 1.20$\pm$0.12 & 2.60 & 5.7 & 10.09$\pm$0.04 \\  
588017703471611920 & ... &  189.9 &  12.4389 &  0.04086  &  10.76 & 383  & $<$1.28  & 2.25 & 6.5 & $<$ 9.97  \\  

\hline
\end{tabular}
\label{preethi_tab}
\end{center}
\medskip
The AGC number refers to the ALFALFA catalog number and exists only for 
detections.  W$_{50}$ is the FWHM of the 21 cm emission line for detections 
and derived from the $i$-band Tully-Fisher for non-detections. S$_{21}$ 
is the integrated 21 cm line flux.  Derivations of the RMS and S/N are 
given in Haynes et al. (2011) and are set to 2.25 mJy
and 6.5 for non-detections (ALFALFA.40 limits). Stellar masses are taken
from the single Sersic fits of Mendel et al. (2014). 
\end{table*}

\begin{figure}
\centerline{\rotatebox{0}{\resizebox{10cm}{!}
{\includegraphics{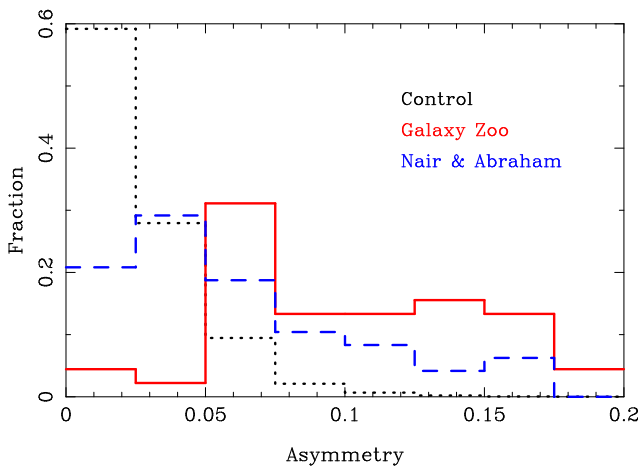}}}}
\caption{\label{asym} The distribution of asymmetry ($R_A$)
parameters in the 45 Galaxy Zoo post-mergers and 48 post-mergers from
Nair \& Abraham (2010) with HI mass measurements, plus the
combined control sample (red, blue dashed and black dotted 
histograms respectively). Whereas the Galaxy
Zoo post-merger
sample has very few mildly asymmetric galaxies ($R_A < 0.05$), the
Nair \& Abraham sample includes galaxies with much more subtle
asymmetries.  Both post-merger samples have significantly higher asymmetries
than the control sample.}
\end{figure}

\begin{figure}
\centerline{\rotatebox{0}{\resizebox{8cm}{!}
{\includegraphics{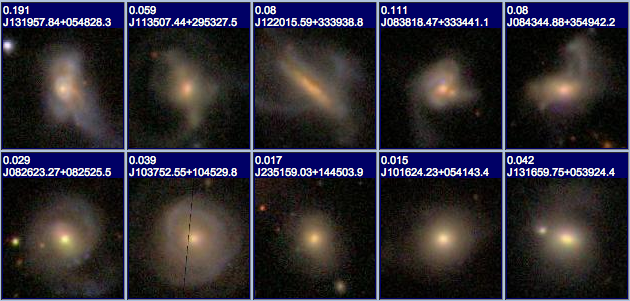}}}}
\caption{\label{asym_images} Examples of highly asymmetric ($R_A > 0.05$)
galaxies in the Galaxy Zoo sample (top row) and smoother ($R_A < 0.05$)
galaxies from the Nair \& Abraham supplementary sample in the lower
row. The value of $R_A$ is given at the top of each panel.}
\end{figure}

In order to expand our post-merger sample, we used the visual
classifications of $\sim$ 14,000 galaxies from Nair \& Abraham (2010).  
The Nair \& Abraham classifications are also based on SDSS images, 
but from the DR4 and with slightly different magnitude and redshift
thresholds from those of Galaxy Zoo: $m_g < 16$ and 0.01 $< z < 0.1$.
However, although the Nair \& Abraham catalog is largely a subset
of the much larger Galaxy Zoo sample, complementary classifications
exist, due to the different approaches taken by classifiers (the statistical
output of non-expert crowd-sourcing vs. a single expert eye).
In particular, the Galaxy Zoo classifications are biased towards
the most highly disturbed mergers, which are obvious from
a brief inspection of the SDSS postage stamp images provided
to the public.  Conversely, the Nair \& Abraham
(2010) catalog additionally identifies more subtle, and often low
surface brightness, features such as faint shells and tidal
tails, as we demonstrate below.

We compile the supplementary sample of post-mergers in two steps.
First, we make use of 411 potential late-stage post-mergers,
galaxies with strong shells and tidal features and close companions
with merger classifications not present in the original Nair \&
Abraham (2010) catalog.  We supplement this with a query of the published
catalog to identify galaxies with an interaction\_flag $>$ 2, but with
the pair\_flag = 0, which yields 77 galaxies.  These lists are not unique.
Both lists are cross-matched with ALFALFA.40, combined, and then duplicates
between the two, and with the Ellison et al. (2013a) sample,
are excluded.  The final Nair \& Abraham post-merger sample
contains 48 galaxies with 
ALFALFA matches, of which 23 have detections and 25 have upper limits
(Table \ref{preethi_tab}).  As for our targeted Arecibo sample, we
list the FWHM ($W_{50}$) of the 21 cm emission line for detections 
and the width derived from the $i$-band Tully-Fisher for non-detections,
as described in the previous section. 
Derivations of the RMS and S/N are given in Haynes et al. (2011) and are 
set to 2.25 mJy and 6.5 for non-detections in the ALFALFA.40 sample.

\subsection{Control Sample}\label{control_sec}

In order to investigate differences between the post-mergers and
the general population, it is necessary to define a control
sample of undisturbed galaxies (defined as having a Galaxy Zoo
merger vote fraction of zero) with no close companion (e.g. Ellison
et al. 2013a,b).  The
control sample is compiled by matching non-merger galaxies
simultaneously in stellar mass, redshift and environment\footnote{Unfortunately,
without precise information on the progenitor morphologies, it is
not possible to account for Hubble type in the matching process.}.
Following our previous works, we compute local environmental densities, 
$\Sigma_n$:

\begin{equation}
\Sigma_n = \frac{n}{\pi d_n^2},
\end{equation}

where $d_n$ is the projected distance in Mpc to the $n^{th}$ nearest
neighbour within $\pm$1000 \kms.  Normalized densities, $\delta_n$,
are computed relative to the median $\Sigma_n$ within a redshift slice
$\pm$ 0.01.  In this study we adopt $n=5$. The tolerance for matching
is 0.005 in redshift, 0.1 dex in stellar mass and 0.1 dex in normalized
local density.  If less than five matches are found, the
tolerance is grown by a further $\Delta z$=0.005 in redshift,
$\Delta$log M$_{\star}$=0.1 dex in stellar mass and  $\Delta \delta_5$=0.1 dex
in normalized local density until the required number of matches is achieved.  

The control matching accounts for various intrinsic dependences and
observational biases.  Perhaps most importantly, it has been shown 
that HI mass correlates well (although with significant scatter)
with stellar mass (Fabello et al. 2011;
Cortese et al. 2011; Catinella et al. 2012).  Looking
for deviations in the typical HI content therefore needs to
take this scaling into account.  Due to the fixed exposure time of
the ALFALFA survey, the HI detection threshold is
strongly distance dependent (Haynes et al. 2011), necessitating the
redshift matching.  Finally, the environmental
matching mitigates biases due to contamination by other galaxies in the
Arecibo beam (although manual rejection of cases with obvious beam
contamination are performed as part of the catalog assembly), as well
as any dependence of gas fraction on group or cluster membership,
or other environmental trends (e.g. Solanes et al. 2001; Kilborn et al. 2009;
Chung et al. 2009; Cortese et al. 2011;
Fabello et al. 2012; Catinella et al. 2013; Hess \& Wilcots 2013).  

\subsection{Sample comparison}

\begin{figure*}
\centerline{\rotatebox{0}{\resizebox{18cm}{!}
{\includegraphics{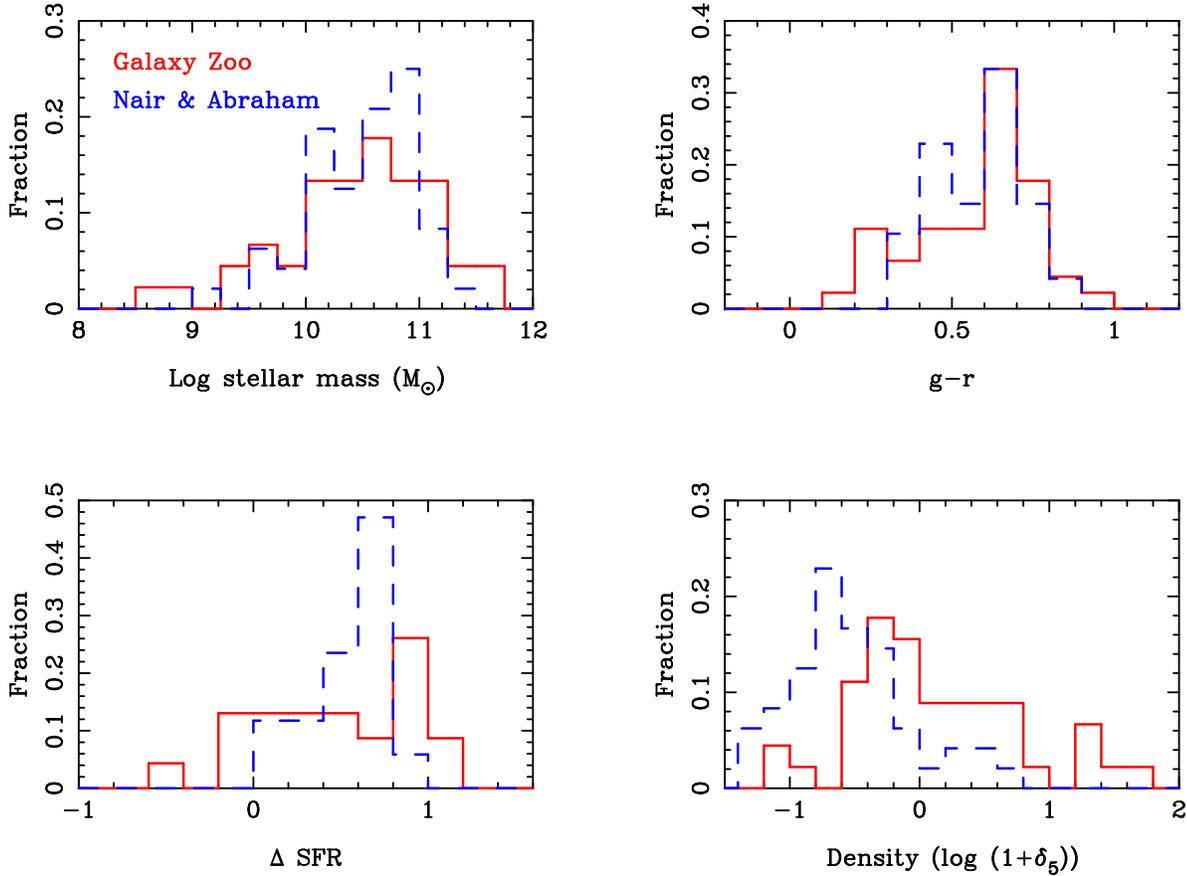}}}}
\caption{\label{comp_pre} A comparison between the properties of
the Galaxy Zoo post-merger sample and the  Nair \& Abraham supplementary
sample (red and blue-dashed histograms respectively) for which we have 
HI measurements.  Top left panel:  Total stellar masses are taken from Mendel
et al. (2014).   Top right panel:  Extinction and k-corrected
Petrosian $g-r$ colours. Bottom left panel:  SFR enhancements. 
Bottom right panel: normalized local densities. }
\end{figure*}

The final post-merger sample used in this paper is the combination
of the Ellison et al. (2013a) Galaxy Zoo sample and the
Nair \& Abraham (2010) supplementary sample, and contains 93 post-mergers
with HI masses (43 galaxies) or limits (50 galaxies).  
A summary of the number of targets in each sub-sample is given in Table
\ref{sample_numbers}.
In Figure \ref{asym} we compare the asymmetries of the Galaxy
Zoo and supplementary Nair \& Abraham (2010) samples that have
ALFALFA.40 matches.  The asymmetries are determined from the
\textsc{gim2d} bulge+disk decompositions of SDSS galaxies performed by
Simard et al. (2011).  In particular, we use the $R_A$ parameter (measured
in the $r$-band),
which measures the fraction of light in the residual image
(after subtraction of the bulge and disk) that remains after
a 180 degree subtraction of itself (see equation 11b
in Simard et al. 2002).   $R_A$ is therefore the 
residual asymmetric component.  

Figure \ref{asym} reveals
that the Galaxy Zoo post-merger sample is dominated by galaxies with
strong asymmetries ($R_A > 0.05$).  Although some highly
asymmetric galaxies do exist in the supplementary 
sample\footnote{Recall that the Nair \& Abraham (2010)
supplementary sample of post-mergers is
checked against the Galaxy Zoo sample and duplicates excluded.
The supplementary sample therefore has had some of its more
asymmetrc galaxies excluded.}, it is clear that it contains
a significant population of galaxies with  $R_A < 0.05$.
For comparison, the asymmetries
of the control galaxies are also shown in Fig. \ref{asym};
60 per cent of the control galaxies have $R_A < 0.02$,
a regime little populated by the post-mergers.
Figure \ref{asym_images} shows some examples of galaxies
with different asymmetries.  In the top row we show five
of the Galaxy Zoo sample with $R_A > 0.05$ and in the
bottom row five of the Nair \& Abraham galaxies with
$R_A < 0.05$.  The value of $R_A$ is given at the
top of each panel.  

\begin{table}
\begin{center}
\caption{The number of post-mergers in each sub-sample}
\begin{tabular}{lcc}
\hline
Selected from:  &  Ellison et al.   & Nair \& Abraham \\
                &  (2013a) / Galaxy Zoo   &  (2010) \\
\hline
Total           & 45  & 48 \\
HI detections   & 20  &  23 \\
Above ALFALFA.40 &  15  & 16\\
90\% completeness &  & \\
\hline
\end{tabular}
\label{sample_numbers}
\end{center}
\end{table}

From simulations of galaxy mergers, it has been shown that the visual
properties of post-mergers depend on timescale, the initial
galactic properties, merger geometry and mass ratio 
(Lotz et al. 2008, 2010a,b; Ji et al 2014).  For example, 
Lotz et al. (2010a) have
demonstrated that higher gas fraction galaxies exhibit longer-lived
tidal features and disturbed morphologies.  Conversely, `dry mergers',
those with low gas fractions, will exhibit lower surface brightness,
less dramatic, and shorter-lived features.  Post-mergers with
lower asymmetries (i.e. smoother morphologies) may therefore be
the result of lower gas fraction mergers, or they may simply be older.
Trends of gas fraction with galaxy asymmetry are discussed later in this paper.

The properties of the two post-merger samples are further explored
in Fig. \ref{comp_pre}.  Recall that some of the post-mergers in the
 Nair \& Abraham (2010)
catalog are also identified by Galaxy Zoo and duplicates are removed from
the supplementary sample.  The supplementary sample
of post-mergers presented here represents only those galaxies
uniquely identified by Nair \& Abraham (2010).  
The top left and right panels of Fig. \ref{comp_pre} show that the 
stellar mass and Petrosian $g-r$ colour distributions are similar
between the two samples.    The bottom left
panel of Fig. \ref{comp_pre} compares, for galaxies classified as
star-forming, the SFR
enhancements, relative to th mass-, redshift and environment-matched
control sample described in Section \ref{control_sec} 
($\Delta$ SFR, see Ellison et al. 2013a,b for more details).
In brief, $\Delta$ SFR compares the SFR of the post-merger with
that of its control sample:

\begin{equation}
\Delta SFR = \log SFR_{PM} - \log SFR_{control},
\end{equation}

such that, for example, a factor of two enhancement in SFR yields a 
$\Delta$ SFR = 0.3.  The median SFR enhancement is 0.41 for the Galaxy Zoo 
postmergers,
but 0.62 for the Nair \& Abraham sample.  Moreover, there are no
galaxies in the latter sample with a negative $\Delta$ SFR (lower
SFRs than the matched control).  However, the \textit{fraction}
of star-forming galaxies is significantly lower in the Nair \& Abraham 
sample: 35 per cent, compared with 52 per cent amongst the Galaxy Zoo
post-mergers. 

There is also a difference in the local environments of the two
samples, as shown in the lower right panel of Fig. \ref{comp_pre},
with post-mergers in the supplementary sample systematically
located in regions of lower galactic density.  However, as
noted above, the procedure for sample selection and removing
duplicates means that a comparison of galaxy properties between the two
samples is not strictly based on physical properties
(an analysis that we perform later in this paper).
Nonetheless, the above statistics and Figs \ref{asym} -- \ref{comp_pre}
demonstrate that the two visually selected samples identify galaxies
with different properties, despite being selected from the same
imaging data.

\subsection{Completeness}

Of the 93 post-mergers presented in this study, 37 have new HI mass
measurements from our Arecibo observations.  It is interesting to
compare the sensitivity of these new observations to that of the
ALFALFA survey.  As described in Haynes et al. (2011), the completeness
of the ALFALFA survey is basically a function of the expected
FWHM of the 21cm line, $W_{50}$.  From equation 4 of Haynes et al. (2011),
we have that the 90\% completeness of code 1 sources (S/N$>$6.5) is given as

\begin{equation}\label{comp_eqn}
 \log S_{21} =
  \begin{cases}
   0.5 \log W_{50} - 1.14,  & \mbox{if }  \log W_{50} < 2.5
  \\
    \log W_{50} - 2.39,  &  \mbox{if } \log W_{50} \ge 2.5
  \end{cases}
\end{equation}

\begin{figure}
\centerline{\rotatebox{0}{\resizebox{10cm}{!}
{\includegraphics{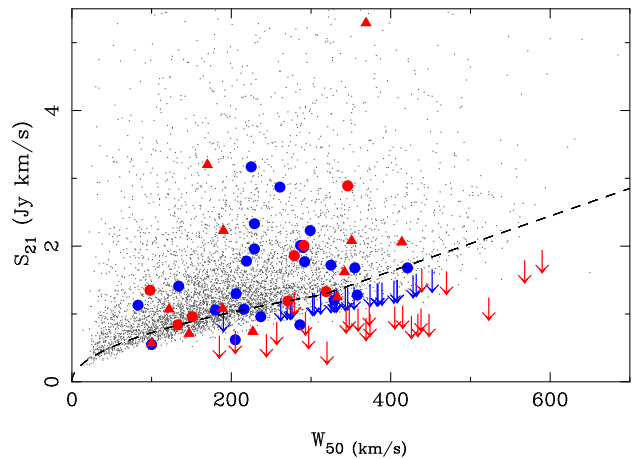}}}}
\caption{\label{complete} The 21 cm flux ($S_{21}$) as a function of the
 FWHM ($W_{50}$).
The black dashed line shows the 90\% completeness of the ALFALFA.40 catalog, as
given by Haynes et al. (2011), see Eqn \ref{comp_eqn}.  ALFALFA.40 (non-merger)
galaxies are shown in grey.  Post-mergers are shown in red and blue for the 
Galaxy Zoo
(Ellison et al. 2013a) and Nair \& Abraham (2010) samples, respectively.
Of the 45 Galaxy Zoo post-mergers, new detections are shown with red
triangles, new limits are shown with red arrows  and matches to ALFALFA.40 
are shown with red circles.
All of the Nair \& Abraham (2010) post-mergers have detections taken
from ALFALFA.40, and are therefore all shown with circles. }
\end{figure}

In Fig \ref{complete}, we show the measured fluxes of ALFALFA.40 code 1 sources, with
the 90\% completeness limit shown.  We show in red the 45 measurements of Galaxy
Zoo post-mergers (8 of which are taken from ALFALFA.40), and in blue we show 
the 48 measurements from ALFALFA.40 of the post-mergers from the Nair \& Abraham
(2010) catalog.  It is clear that some of our new observations are deeper than
the completeness limit of ALFALFA, which is to be expected from our observation strategy.
In the following analysis, we will frequently consider only those post-mergers that
lie above the completeness limit, in order to make fair statistical comparisons
between the gas content of the post-mergers and isolated (non-merger)
galaxies.  However,
for the rest of this analysis, we will combine the supplementary Nair \& Abraham
sample with the Galaxy Zoo post-merger sample.  We hereafter refer to this simply
as `the post-merger' sample; it contains 93 galaxies, of which 43 have HI detections
and 31 are above the ALFALFA.40 90\% completeness limit (Table \ref{sample_numbers}). 
Although we have shown that
the properties of these samples differ, their properties overlap and the
distinction between them has no clean physical boundary.  Later in this work,
we will investigate how gas fractions depend on galaxy properties within this
combined sample.

\section{Results}

\subsection{Gas fractions}

\begin{figure}
\centerline{\rotatebox{0}{\resizebox{10cm}{!}
{\includegraphics{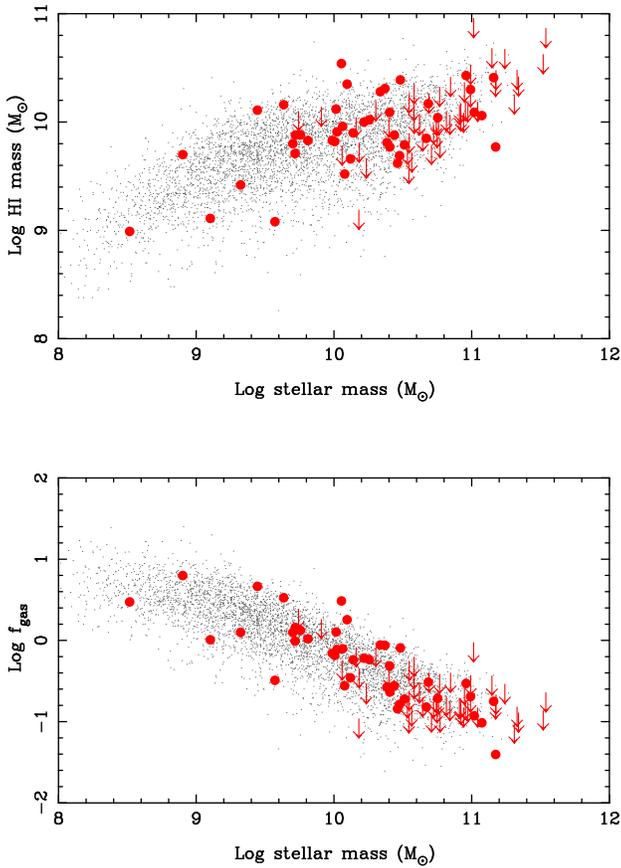}}}}
\caption{\label{mass} The HI mass (top panel) and HI gas fraction (bottom
panel) as a function of stellar mass.  Red points (and upper limits)
represent measurements for the post-merger sample and small grey points are 
non-merger galaxies extracted from the ALFALFA.40 data release (see text for details).  
  }
\end{figure}

The HI gas fraction (hereafter simply `gas fraction') is defined
in this work as $f_{gas} =$ M$_{\rm HI}$ / M$_{\star}$\footnote{In theoretical
works, it is more common to define $f_{gas} =$ M$_{\rm HI}$ / M$_{total}$,
but we adopt the more frequently used definition in observational studies.}.
In Figure \ref{mass} we present the HI masses and gas fractions for the
full post-merger sample.  For comparison, we also show non-merger
galaxies detected in the ALFALFA.40 sample.  The isolated (non-merger) 
nature of 
these galaxies is determined by requiring that the Galaxy Zoo merger vote 
fraction be zero, and we further require that the ALFALFA ocode=I and Ngal=1, 
indicating that only one  SDSS source has been matched.  For clarity, 
only ALFALFA detections (not upper limits)
are shown.    Figure \ref{mass} shows that the galaxies in our post-merger sample
are biased towards the high mass end of the galaxy distribution
for which ALFALFA HI masses exist.     Figure \ref{mass} also shows that
 some of our own Arecibo observations are deeper than the typical 
ALFALFA.40 detections (see also Figure \ref{complete}), whose effective 
integration times are only 40 seconds (recall that ALFALFA upper limits are 
not shown to avoid crowding).  

\begin{figure}
\centerline{\rotatebox{0}{\resizebox{10cm}{!}
{\includegraphics{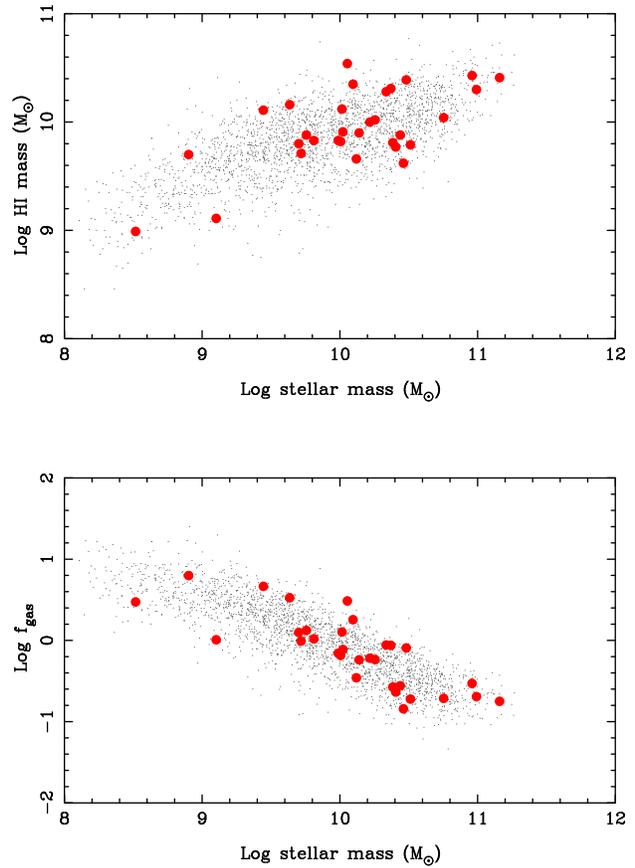}}}}
\caption{\label{mass_complete} As for Fig. \ref{mass}, but only showing
galaxies with HI detections that are above the ALFALFA.40 90\% completeness limit. }
\end{figure}

In order to make a more fair comparison between the post-mergers and
ALFALFA.40, Fig. \ref{mass_complete} shows the same data as in 
Fig. \ref{mass} but with only those
galaxies that pass the 90\% completeness threshold included.  
In order to test statistically whether post-mergers exhibit
distinct gas fractions from isolated galaxies, we will define a galactic
`gas fraction offset', $\Delta$ \fgas, which quantifies a gas 
fraction relative to isolated (control) galaxies with otherwise similar 
properties.  The $\Delta f_{gas}$ metric is very similar in spirit
(although opposite in sign) to the widely used HI deficiency
(e.g. Haynes, Giovanelli \& Chincarini 1984).  The application of such 
`offset' techniques has been previously used to effectively quantify 
differences in colour, SFR (e.g. Fig. \ref{comp_pre}),  metallicity and 
active galactic nuclei (AGN) frequency between galaxy pairs 
and non-mergers  (e.g. Patton et al. 2011, 2013; Ellison et al. 2010, 
2011b; Scudder et al. 2012; Satyapal et al. 2014).

\subsection{Gas fraction offsets}

The basic procedure for control matching is described in Section
\ref{control_sec}, although we now additionally require that
both the post-merger, and galaxies in the control `pool'
lie above the ALFALFA  completeness threshold, as defined in equation 
\ref{comp_eqn}.  There are 31 post-mergers that fulfill this criterion
(see Table \ref{sample_numbers}) and can hence have robust gas
fraction offsets quantified. Mergers 
that lie below the completeness threshold will appear to have spuriously
low HI masses for their stellar mass, due to the lack of ALFALFA galaxies
in this regime.      As described in Section \ref{control_sec},
the control sample is assembled by simultaneously
matching in stellar mass, redshift and local environment all ALFALFA
galaxies that are also above the 90\% completeness threshold.  By definition,
this is the majority, but not all, of the published ALFALFA.40 sample.
As described in Section \ref{control_sec}, we require a minimum of 5
control galaxies for each post-merger; if this minimum is not reached,
the matching tolerances are grown.  In practice, there are typically
10 -- 30 controls for each post-merger.

From the matched control sample it is then possible to calculate a 
gas fraction offset defined as 

\begin{equation}
\Delta {f_{gas}} = \log f_{\rm{gas,PM}} - \log f_{\rm{gas,control}},
\end{equation}

where the control value is the median of the matched control galaxy
gas fractions for a given post-merger.
This calculation is only made for galaxies with detections above
the ALFALFA.40 90 per cent completeness threshold.
A value of $\Delta f_{gas}=0$ therefore indicates that a galaxy
has an HI gas content that is typical for its mass, redshift and
local density.  In contrast, positive and negative
values of $\Delta f_{gas}$ indicate gas-rich and gas-poor galaxies 
respectively.  
We can also calculate gas fraction offsets for isolated ALFALFA
galaxies, by again comparing their gas fractions with other isolated
galaxies at the same mass, redshift and local density.

\begin{figure}
\centerline{\rotatebox{0}{\resizebox{10cm}{!}
{\includegraphics{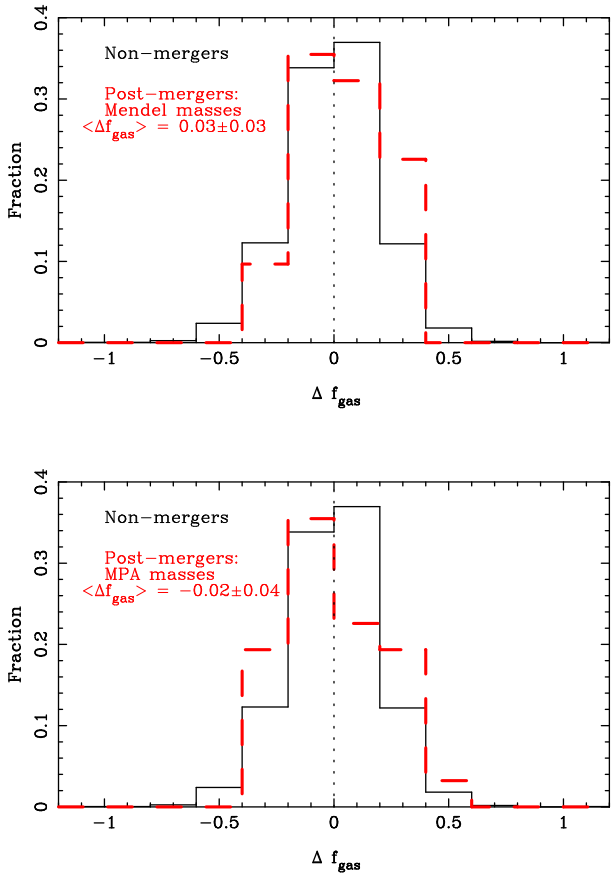}}}}
\caption{\label{dfgas} Distribution of gas fraction offsets.  In both
panels, the solid black histogram shows the distribution of $\Delta f_{gas}$ 
for isolated ALFALFA galaxies and the red dashed histograms indicate
the post-merger sample.  In the top panel, the stellar masses of post-merger
galaxies are taken from the Mendel et al. (2014) catalog.  In the lower
panel, stellar masses are taken from the MPA catalog (Kauffmann et al. 2003).  }
\end{figure}

The top panel of Figure \ref{dfgas} shows the distribution of $\Delta
f_{gas}$ amongst isolated ALFALFA galaxies (black histogram) and
the post-merger sample (red dashed histogram).  For the 
post-merger galaxies that are above the ALFALFA.40 completeness 
threshold the distribution of gas fraction offsets is very similar to
the values observed in isolated galaxies.  The mean gas fraction
offset in the post-mergers is consistent with no merger-induced 
change in the atomic gas fraction: $\Delta f_{gas} = 0.03 \pm 0.03$.

Since the gas fraction combines both HI masses and stellar masses
by definition, it is germane to consider whether the gas fraction
results depend sensitively on the derivation of M$_{\star}$.
 We have adopted the
single Sersic fits from Mendel et al. (2014), which include the
numerous improvements to SDSS photometry detailed in Simard et al. (2011).
However, there are two possible concerns with these masses that
are relevant to the study of post-mergers.  First,
a single Sersic fit does not account for flux in the asymmetric components
of the light, which might be a significant fraction in post-mergers.
Second, the Mendel masses do not include star formation histories with
bursts.  Although smooth star formation histories are a good average
representation of the general galaxy population, this is not likely to be
the case for mergers, which are known to exhibit bursts of star
formation following close passages and at coalescence.
For comparison, we therefore repeat the $\Delta f_{gas}$ analysis with
masses derived by Kauffmann et al. (2003) (hereafter `MPA masses')
that are based on total
photometry and include bursty star formation histories.  The comparison
between values of $\Delta f_{gas}$ for the ALFALFA.40 sample
and post-mergers derived from matching MPA masses, rather than Mendel
masses, is shown in the lower panel of Fig. \ref{dfgas}.  
Although the individual values of $\Delta f_{gas}$ in the post-mergers 
are slightly different, the distribution of values in the post-merger 
sample remains very similar to the isolated sample.  The mean
gas fraction offset is also still consistent with the matched control
sample: $\Delta f_{gas} = -0.02 \pm 0.04$.

\begin{figure}
\centerline{\rotatebox{0}{\resizebox{10cm}{!}
{\includegraphics{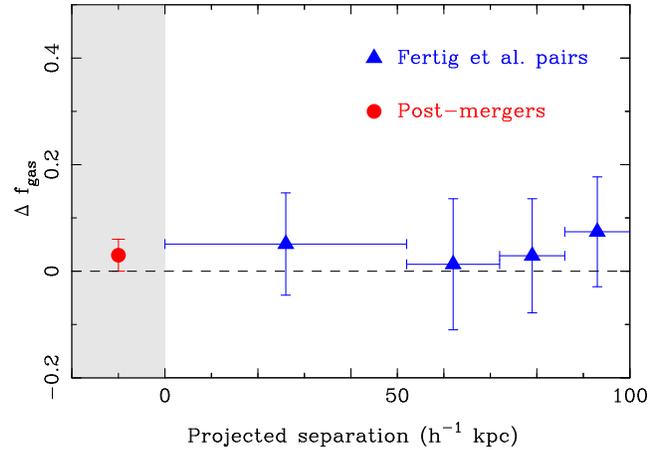}}}}
\caption{\label{dfgas_rp} Distribution of mean gas fraction offsets as
a function of projected separation for the pairs sample of
Fertig et al. (2015, blue triangles) and the post-mergers 
(red circle located in the grey shaded box) presented in this
paper.  There is no indication
for significant gas consumption at any stage during the merger.   }
\end{figure}

Fertig et al. (2015) have recently used similar methods to ours to
study the gas fraction offsets in galaxy pairs.  The only notable difference
in their analysis (apart from the selection of pairs, rather than
post-mergers) is that their control sample is composed of wide pairs, rather
than truly isolated galaxies.  The use of wide pairs as controls
has been previously used in the literature (e.g. Scott
\& Kaviraj 2014) and Fertig et al. (2015) show that this is a robust
approach for gas fractions.  Therefore, although Fertig et al. (2015)
use wide pairs, and this work uses isolated non-mergers, for the control
sample, the methodology is otherwise identical and the results should be 
comparable.

Fertig et al. (2015) found no difference between
the gas fractions in the close and wide pairs, indicating that consumption
is not significant in the pre-merger phase of the interaction. 
In  Fig. \ref{dfgas_rp} we combine the results of Fertig et al. (2015)
with our post-mergers, in order to provide a complete view of HI gas
consumption during the merger process. The blue points show the
gas fraction offsets as a function of projected separation in the
Fertig et al. (2015) pairs sample.  The filled red point is the
mean $\Delta f_{gas}$ in the combined post-merger sample (Fig.
\ref{dfgas}).  None of the observed data points are located at
$\Delta f_{gas} < 0$, indicating a lack of measured gas consumption 
throughout the merger stages probed by our samples.

\subsection{Detection fractions}

In order to make a fair comparison between the post-merger sample
and our control sample (drawn from ALFALFA.40), it was necessary to
impose a consistent detection threshold for both samples.  Although
this leads to a robust and fair comparison, the analysis of
$\Delta f_{gas}$ includes only  one third of our sample
galaxies.  Therefore, whilst we can state that, for galaxies detected
in HI above the ALFALFA threshold the typical HI gas fraction is
consistent with isolated galaxies, we have not yet investigated
the nature of the rest of the sample.  It is possible, for example,
that whilst some galaxies possess  `normal' gas fractions,
the majority of post-mergers are actually depleted, and that this
depletion is the main reason they are below our detection limit.

In the absence of a deeper control sample, it is not possible 
to quantify
the depletion in either the detections that are below the ALFALFA.40 
threshold (of which there are 12), nor in the non-detections.
However, we can compute the detection fractions of the post-mergers and
compare it to that of the control, in order to judge whether the
former are more frequently gas-poor.  For this calculation, we
re-make the control sample, identifying the single best match (minimized
deviation in stellar mass, redshift and local density) for
each post-merger.  This adjustment is necessary due to the different number of
control galaxies per post-merger permitted by the original procedure.
The detection fraction is defined as the fraction of galaxies above
the ALFALFA detection threshold (equation \ref{comp_eqn}), such that this
calculation is not affected by the higher sensitivity of some of our
Arecibo observations.
We find that the detection fraction of post-mergers is almost
double that of the control (32$\pm$6 per cent versus 17$\pm$4 per cent).  The
result is not sensitive to the number of controls required in the
matching process (i.e. similar results are obtained with
5 matches).

The high detection fractions measured amongst the 
full post-merger sample
again support the lack of a dominant population of gas depleted galaxies.
On the contrary, the factor of two enhancement in the post-merger HI
detection fraction may indicate that either the post-merger
galaxies have a higher tendency to contain \textit{some} gas, or
that their gas fractions have been preferentially
boosted above the nominal ALFALFA completeness threshold.  These
possibilities can be
reconciled with the gas fraction result (which yielded an average
post-merger $\Delta f_{gas}$ consistent with zero) by recalling that
ALFALFA is a shallow survey that itself consists of relatively gas-rich
members of the galaxy population.  Our measurement of gas fraction offsets
may therefore be an under-estimate of the true values\footnote{The same
caveat is applied to measurements of other properties of mergers investigated
with matched control samples, due to the intrinsic nature of detection
thresholds. For example, Scudder et al. (2012)  caution that the SFR enhancements 
in their sample of galaxy pairs may be under estimates if their control
sample contains (isolated) galaxies which themselves have preferentially
elevated SFRs that boost them across the detection threshold.}. 
Therefore, not only is there no evidence for gas consumption in the post-mergers based on 
direct comparison of the observed gas fractions in the control and post-merger sample
(Fig \ref{dfgas}, but our analysis of the HI mass detection fractions hint 
that post-mergers may actually be somewhat gas-rich.
This possibility is explored further below in Section \ref{discussion}.

\subsection{Dependence of Gas Fraction Offset on Asymmetry}

\begin{figure}
\centerline{\rotatebox{0}{\resizebox{10cm}{!}
{\includegraphics{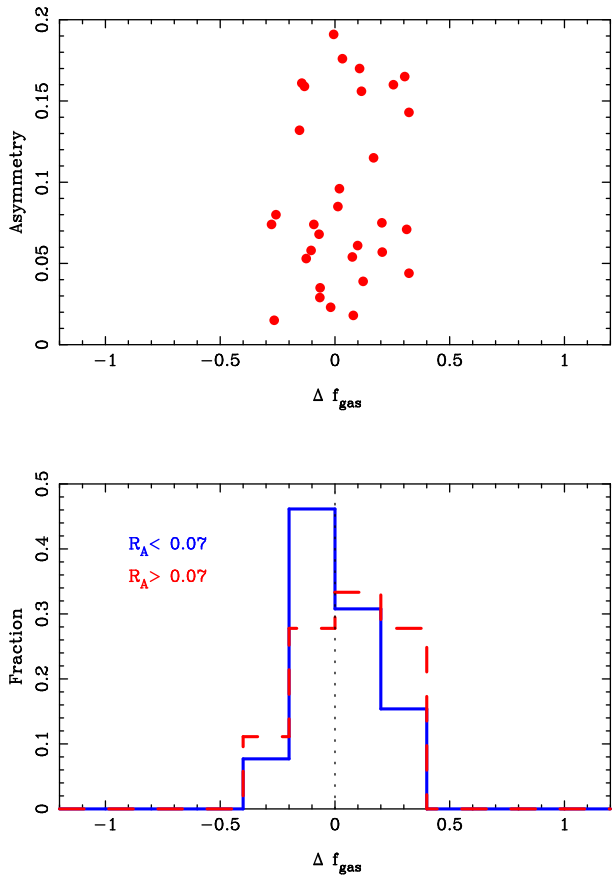}}}}
\caption{\label{dfgas_asym} Top panel: Gas fraction offsets as a function
of galaxy asymmetry for the post-merger sample.  
Bottom panel: Distribution of $\Delta f_{gas}$ for 
post-mergers split by asymmetry  at $R_A =0.07$;
the red and blue histograms show high and low asymmetry galaxies respectively.}
\end{figure}

It is well known that post-mergers are most asymmetric soon
after coalescence, and that tidal features will fade with
time thereafter (e.g. Lotz et al. 2008), so that mergers selected from
shallow imaging surveys will be biased towards recently coalesced galaxies
(Ji et al. 2014).  In addition to the effects of merger timing, the
nature of the progenitor galaxies can affect the resulting asymmetries.
For example, the simulations of Lotz et al. 
(2010a) show that smoother merger remnants result from more gas-poor mergers.

 In Fig. \ref{asym} we showed that
there is a range of asymmetries in our post-merger sample,
with a bias towards the highest values in the Galaxy Zoo
sample from Ellison et al. (2013a).  Indeed, in Ellison et
al. (2013a) we argued that these galaxies were only recently
coalesced based on their low gas phase metallicities.   It
is therefore interesting to investigate whether there is
a trend of gas consumption with asymmetry, as might be expected
if the least disturbed post-mergers are older.

In the top panel of Fig. \ref{dfgas_asym} we plot $\Delta f_{gas}$ 
as a function of $r$-band asymmetry for the ALFALFA.40 complete
post-merger sample.  There is no obvious trend between asymmetry and
gas fraction offset.  In the lower panel
we plot the distribution of $\Delta f_{gas}$ for high
($R_A \ge 0.07$, red histogram) and low ($R_A < 0.07$, blue histogram)
asymmetry post-mergers, relative to the median asymmetry of the sample.
Based on the results of a Kolmogorov-Smirnov (KS) 
test, there is no statistical difference that
the two asymmetry samples have different distributions of $\Delta f_{gas}$.

An important caveat to the application of asymmetry metrics to mergers
is their relative insensitivity to extended low surface brightness
features.  Adapting classic asymmetry indices to be more sensitive to
such low surface brightness features, particularly at large galactocentric 
radii, may help reveal a closer connection with changes manifest in galactic
stellar and gaseous properties (e.g. Lelli et al. 2014; Pawlik et al.,
in prep.).

\section{Discussion}\label{discussion}

One of the main goals of the work presented here is to tackle
the question of whether the merger process significantly depletes
the gas reservoir of the remnant.  We have found no evidence of this in 
the measured HI gas fractions of our sample of post-merger galaxies.
Specifically, based on HI detections in 31 galaxies for which 
we can identify a fair
control sample from ALFALFA, we find that post-mergers have gas fractions 
that are indistinguishable from equivalent isolated galaxies.  However,
due to the shallow nature of the ALFALFA.40 survey, the sample from 
which the controls are drawn is dominated by relatively gas-rich galaxies.  
Indeed, considering the full sample of 93 post-mergers, we find an HI
detection fraction that is double that of the control sample of
isolated galaxies.   Therefore, contrary to finding evidence for gas
consumption, our analysis of HI gas fraction enhancements 
suggests that post-mergers might be relatively 
gas-rich.  A number of other observational studies have drawn
similar conclusions (Braine
\& Combes 1993; Bettoni et al. 2001; Huchtmeier et al. 2008;
Stark et al. 2013; Goncalves et al. 2014). Previous simulations
have proposed that mergers might renew their cold gas reservoirs
through the cooling of hot halo gas (Moster et al. 2011; Tonnesen
\& Cen 2012).  Indeed, recent cosmological simulations measured
increases in the HI gas fraction of $\sim$ 20 per cent within
a few hundred Myr after a merger (Rafieferantsoa et al. 2014).

\begin{figure}
\centerline{\rotatebox{0}{\resizebox{10cm}{!}
{\includegraphics{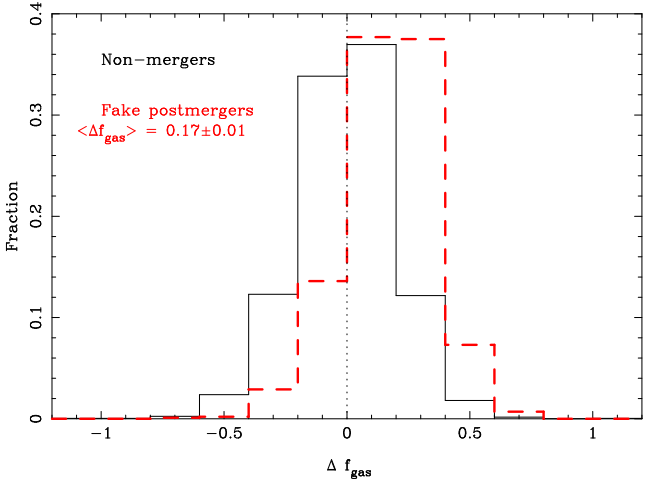}}}}
\caption{\label{fake} Distribution of gas fraction offsets for isolated ALFALFA
galaxies (solid black histogram) and 1000 fake post-mergers (red dashed
histogram), constructed through random draws from the ALFALFA.40 sample. 
Due to the declining gas fraction as a function of stellar mass, matching
in stellar mass leads to a shift towards positive values of $\Delta f_{gas}$. }
\end{figure}

However, despite the theoretical predictions for enhanced gas fractions, 
there may be a more mundane way to create high gas fractions in recent
post-mergers, originating from the declining gas fractions with increasing stellar
mass, as shown in Figure \ref{mass} (and see also Catinella et al. 2012
and references therein).
The simple addition of two randomly chosen galaxies in such a distribution
will yield relatively high gas masses for a given stellar mass.
We demonstrate this effect in Figure \ref{fake} with
a simple simulation.    We draw 2000 galaxies at random from
ALFALFA.40, pairing them up by adding their stellar masses and gas masses
to yield `measurements' of $f_{gas}$ for 1000 fake post-mergers.  
For each of these fake 
post-mergers, we assemble a control sample matched in stellar mass and 
determine $\Delta f_{gas}$ as described above\footnote{For the real
post-mergers, we also match in redshift and environment, but this is
not possible in this simulation.}.  For reference, Figure \ref{fake}
again shows (black histogram) the distribution of $\Delta f_{gas}$ for all 
ALFALFA.40 galaxies above the 90\% completeness limit that also appears in the 
two panels of Figure \ref{dfgas}.  The red dashed histogram shows the
distribution of the 1000 fake post-mergers, that are seen to be offset
towards positive values with a mean $\Delta f_{gas} = 0.17 \pm 0.01$. 
This is a direct consequence of the combination of matching a control 
sample in stellar mass and the declining gas fractions with that
same quantity.  

Interestingly, the histograms shown in Fig. \ref{fake} are qualitatively
similar to the results of Rafieferantsoa et al. (2014), who compare
the gas fractions of recently merged galaxies to non-mergers in
a cosmological simulation and find a 20 per cent increase in the former.
We stress that the gas fraction enhancement seen due to this `summation'
effect depends on the gas fractions of the progenitor galaxies, and
also on the distribution of gas fractions in their controls.  If
pre-merger galaxies have initially higher gas fractions, the resulting $\Delta
f_{gas}$ of the post-merger will also be shifted to more positive values.
Our simulation of fake post-mergers is therefore illustrative of the
qualitative offset towards enhanced gas fractions that might be
expected due to stellar mass matching.

The elevated HI detection fraction (by a factor of two) in our post-mergers
compared with isolated controls of the same stellar mass may plausibly be
explained by this effect.  However, we see no net shift towards positive $\Delta
f_{gas}$ (Fig \ref{dfgas}) as indicated might be expected by Fig. \ref{fake}.  
As proposed above, the consistency between post-merger and
control gas fraction offsets may be due to a tendency for the controls
themselves to be relatively gas-rich, which artificially reduces the
magnitude of the measured gas fraction enhancement.  An alternative explanation 
for the lack of gas fraction enhancement in the observations is
that prompt consumption of the initial HI excess has already occured in the
post-mergers due to high rates of star formation.  In order to
quantify the expected amount of gas consumption in the post-mergers
phase, we turn now to a suite of binary merger simulations.

\subsection{Timescales for Gas Depletion}

Fig. \ref{fake} demonstrates that, in a stellar mass
matched sample, the trend of declining \fgas\ with M$_{\star}$ leads
to a shift towards high gas fraction offsets when two galaxies are combined.
Moreover, accretion of new cold gas is predicted
by some merger simulations (e.g. Moster et al. 2011; Tonnesen
\& Cen 2012), and also might be expected in generic scenarios
of mixing a cool and ionized medium (Marinacci et al. 2010; Marasco,
Fraternali \& Binney 2012). However, Fig. \ref{dfgas} shows that 
there is, on average, no enhancement in the gas fractions in the
observed post-mergers. In this section, we explore whether the
contrast between Figs. \ref{dfgas} and \ref{fake} is actually an
indication of gas consumption.

A detailed assessment of gas consumption and recycling
requires sophisticated hydrodynamical simulations that can adequately
track the gas in its different phases (e.g. Dav\'e et al. 2013), 
which is beyond the scope of
this paper (but see Rafieferantsoa et al. 2014).  
However, we can make a rough assessment for a simplified
case, based on the suite of merger simulations presented in Patton
et al. (2013) using the model for radiative gas cooling, star formation, 
and feedback as summarized in Torrey et al. (2012).  Such simulations 
will also provide an approximate
timescale over which galaxies will return to their nominal scaling
relations.

The suite of binary mergers includes 75 orbital variations of a
merger with mass ratio $M_{\star,1} / M_{\star,2} = 1.4 \times 10^{10}$ M$_{\odot}$/
$5.7 \times 10^9$ M$_{\odot}$ = 2.5, which
represents the typical pairing in the SDSS sample of Patton et al.
(2013).   Two isolated galaxies, identical to each of
the galaxies in the merger, which act as controls, are also simulated.    
By comparing the integrated SFR in mergers during the course of the 
simulation (which runs for a total
of 6.5 Gyr), with the isolated controls, we can determine
the typical additional stellar mass created in the merger
(compared to the baseline stellar mass created in the controls),
$\delta$M. Although this simulation does not track the various phases of gas, 
we make the simplifying  assumption that 
star formation is intially fed from gas that originated as HI.
This assumption leads to a conservative upper limit on
the gas consumption.

\begin{figure}
\centerline{\rotatebox{0}{\resizebox{10cm}{!}
{\includegraphics{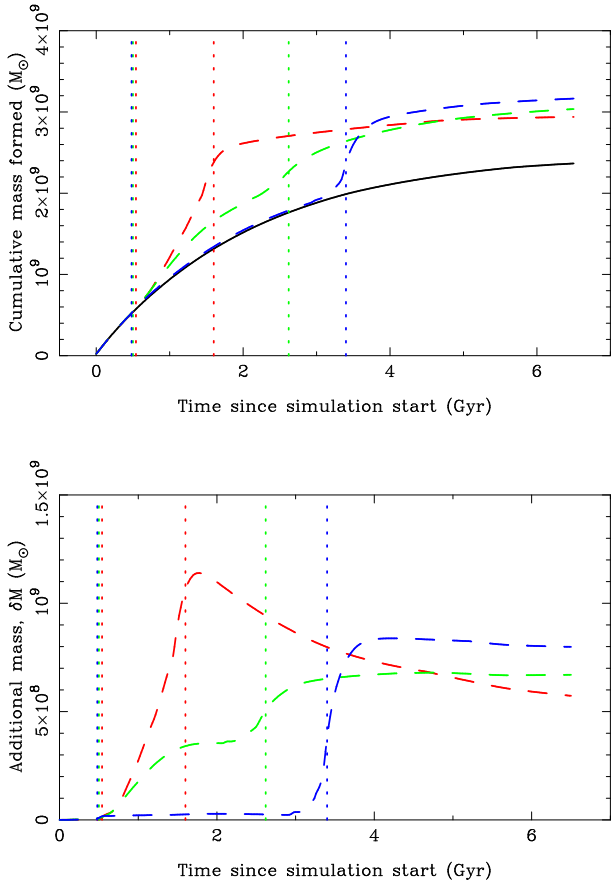}}}}
\caption{\label{3tracks}  Top panel:  Cumulative stellar mass formed
in three example merger simulations from our suite of 75 orbital
combinations of a 2.5:1 stellar mass ratio merger.  The coloured
dashed lines show the combined stellar masses made in both galaxies
througout the 6.5 Gyr simulation.  Red, green and blue lines show
examples from the $e, f$ and $k$ orientations respectively.
The first (left-most) set of vertical
dotted lines indicates the time of first peri-centre.  The second (right-most)
set of vertical dotted lines shows the time of coalescence, which is sensitive 
to the orbital parameters.  The black line
shows the steady evolution of the isolated control galaxies.
Bottom panel:  The difference in stellar mass ($\delta M$) accumulated
by the mergers relative to the control galaxies.  }
\end{figure}

In Fig. \ref{3tracks} we plot the accumulation of stellar mass
throughout the 6.5 Gyr simulation for 3 out of 75 of our orbital
variations, which include combinations of five eccentricities
and five impact parameters for 3 relative disk inclinations 
(Patton et al. 2013; Moreno et al. 2015).  The red, green and blue 
lines show examples from the
\textit{e, f} and $k$ suites, which have relative disk orientations of 60,
90 and 180 degrees respectively (Robertson et al. 2006).  
The coloured dashed lines in the top panel shows the
accumulated stellar mass in both galaxies in the merger, and the
black line shows the mass build-up for identical galaxies evolved
in isolation.  The first (left-most) set of vertical
dotted lines indicates the time of first peri-centre; this time is very
similar for each of the 3 orbits.  The second (right-most)
set of vertical dotted lines shows the time of coalescence, which is more
sensitive to the orbital parameters.  The example $e$ simulation (red
line) is strongly affected by the first pericentric passage and significantly
increases its stellar mass through triggered star formation after
this time.  However, it is little affected by coalescence.  The opposite
is true for the example $k$ merger.  Its anti-aligned disk means that
it is little affected by the pericentric passage, but experiences a
larger star burst at coalescence.  The intermediate example, $f$, has
modest star formation triggered at both peri-centre and coalescence
(see Moreno et al. 2015 for a more detailed analysis of orbital
and spatial dependences of triggered star formation).
Despite these very different histories, the total accumulated stellar
mass is very similar for these three examples.  This general tendency
for small pericentre starbursts to be followed by large bursts at coalescence,
and vice-versa, was also observed in the large suite of merger simulations
run by Di Matteo et al. (2007).  In the lower panel of
Fig. \ref{3tracks}, we calculate the difference between the accumulated
stellar mass in the mergers and that of the control galaxies.  Again,
the time at which the mass is formed is evidently dependent on the geometry,
but converges to a similar value of $\sim 7 \times 10^8$ M$_{\odot}$ by
the end of the simulation.  Note that $\delta M$ does not change much after
500 Myrs post-coalescence, indicating that even if there is a star burst
at the time of the final merger, it is short-lived and gas consumption
does not increase much further beyond the rate expected for the isolated
galaxy.  The range of $\delta M$ (after 6.5 Gyr) amongst the full
suite of 75 orbits spans $3 \times 10^8$ to $1 \times 10^9$ M$_{\odot}$,
again, a relatively narrow range considering the different burst histories.

Having established a typical value of $\delta M = 7 \times 10^8$ 
M$_{\odot}$, we assign 
HI masses ($M_{HI,1},  M_{HI,2}$) to the galaxies by determining the best fit
solution to the ALFALFA.40 data points shown in Fig. \ref{mass_complete}:

\begin{equation}
\log M_{HI} = 0.366 \log M_{\star} + 6.231.
\end{equation}

\noindent The gas fraction of the post-merger is therefore given as

%\begin{equation}\label{sim1}
\begin{multline}\label{sim1}
\log f_{gas,PM} = [\log(M_{HI,1} + M_{HI,2} - 2\delta M)] - \\
 [\log(M_{\star,1} + M_{\star,2} + 2\delta M)]  .
\end{multline}
%\end{equation}

\noindent The control galaxy gas fraction $f_{gas,c}$ is determined from 

\begin{equation}\label{sim2}
\log f_{gas,c} = \log(M_{HI,c}) - [\log(M_{\star,1} + M_{\star,2})].
\end{equation}

\noindent where $M_{HI,c}$ is again determined from the best fit ALFALFA.40
data for a stellar mass $\log(M_{\star,1} + M_{\star,2})$.  In the
case of no star formation (i.e. akin to the simulation of random
pairings shown in the lower panel of Fig \ref{dfgas}) we find a mean
$\Delta f_{gas} = 0.17$.  Inserting the appropriate numbers
into equations \ref{sim1} and \ref{sim2}, we find $\Delta f_{gas} = 0.11$
where star formation is depleting gas over the course of the simulation.
Therefore, for the typical pair represented by our merger suite,
the expected decrease in $\Delta f_{gas}$ even several Gyr after coalescence
is very modest, only 0.06 dex.  For comparison, Fertig et al. (2015) used
the same set of simulations to estimate an expected decrease in $\Delta$
\fgas\ of 0.04 dex in the close pairs phase of the merger.  

Although based on simplified approximations, the
results of our simulations lead to similar conclusions to the
more sophisticated treatment of gas fractions in mergers by Rafieferantsoa
et al. (2014), namely that galaxy mergers may  yield a medium-to-long
term shift towards high gas fractions in their remnants, before
a modest decline.
The results of these simulations indicate that the contrast between the
expected shift to positive $\Delta f_{gas}$ from the summation of galaxies 
(Fig. \ref{fake}) and the lack of a shift in the observed post-mergers
(Fig. \ref{dfgas}) is not due to gas consumption.  Instead, we favour
a scenario in which the matching of control galaxies from the ALFALFA survey
is biased towards gas-rich galaxies, which reduces the gas fraction
enhancement.  However, increased gas fractions (or gas presence)
are nonetheless implied from
the high HI detection fractions in the post-mergers.

\section{Conclusions and closing remarks}

HI mass measurements of 93 post-merger galaxies, visually selected from 
the Sloan Digital Sky Survey, are presented, including new Arecibo
observations for 37 of the sample galaxies.  The remaining HI masses
are taken from the ALFALFA.40 sample (Haynes et al. 2011).  
Through a carefully controlled
matching procedure with isolated (non-merger)
galaxies selected from the ALFALFA.40
data release, and taking into account differences in completeness thresholds,
we calculate the gas fraction
offset, $\Delta f_{gas}$, which quantifies the relative gas richness
of each post-merger.  We also calculate the gas fraction offset for all
(non-merger) ALFALFA.40 galaxies through the same matching procedure.
Finally, we interpret our gas fraction results with the aid of
a suite of 75 binary merger simulations.

\medskip

Our conclusions are as follows:

\begin{enumerate}

\item The gas fractions in post-merger galaxies that 
are above the ALFALFA.40 completeness threshold are consistent
with those measured in a mass-, redshift- and environment-matched
control sample (Fig. \ref{dfgas}).  Combined with comparable
measurements in galaxy pairs (Fertig et al. 2015), we find
that the HI gas fractions in interacting galaxies
are consistent with their controls throughout
the merger sequence (Fig. \ref{dfgas_rp}), with no evidence for
significant consumption of the neutral gas reservoir.

\item  There is no trend between the gas fraction enhancement and 
galaxy asymmetry (Fig. \ref{dfgas_asym}).

\item The HI detection fraction in the full post-merger sample
is approximately double the detection
fraction of the isolated, matched control galaxies.  Therefore,
in addition to finding a lack of gas consumption, our results 
suggest that
post-mergers may be, on average, more gas rich than isolated
galaxies.

\item Although high gas fractions in post-mergers have been predicted
in several simulations (Moster et al. 2011; Tonnesen \& Cen 2012;
Bekki 2014; Rafieferantsoa et al. 2014), we demonstrate that this
enhancement is expected even in the absence of hot halo gas cooling,
or the accretion of external gas.
By drawing random pairs of galaxies from the isolated galaxies
in the ALFALFA.40 sample we show that high gas fractions may arise
naturally from the pairing of two galaxies drawn at random from the
ALFALFA.40 detections, due to the declining \fgas\ with increasing
stellar mass of the underlying population (Fig. \ref{fake}).  
Indeed, the results of this test indicate that for a stellar mass
matched sample we might expect to see a shift towards positive $\Delta$
\fgas\ in post-mergers. 

\item    Based on a suite of binary merger simulations, we compute a conservative
upper limit for HI gas consumption for a galaxy interaction that is representative
of the SDSS pairs sample (Fig \ref{3tracks}).  
We find that, even several Gyrs after coalescence, the impact
of star formation on gas consumption is small and that
$\Delta f_{gas}$ is expected to change very little, decreasing by only 0.06 dex
(although the exact consumption is likely to depend on mass ratios,
orbital geometries etc.).
Gas consumption therefore does not appear to explain the lack
of gas fraction enhancement in the observations.  Instead, we
propose that our gas fraction offsets may be lower limits, due to
the relatively gas-rich nature of the ALFALFA control sample.

\end{enumerate}
\textit{We conclude that there is no evidence for significant gas consumption
in our post-merger observations; on the contrary, the post-mergers 
may actually be somewhat gas-rich.}

\medskip

One remaining uncertainty in our investigation of the evolution of
gas content in mergers concerns the molecular gas content.  
Measurement of the molecular component is not only necessary
for obtaining a complete census of the galactic gas, but it is
also critical for understanding the mode in which star formation is
progressing.  Highly star-forming galaxies at both high redshift and
locally (e.g. the ULIRGs) are offset on the Kennicutt-Schmidt relation
(Daddi et al. 2010a; Genzel et al. 2010).  Mergers have been predicted to
also inhabit this offset sequence that exhibits ten times higher SFR
at fixed gas (total) mass surface density (Renaud et al. 2014).  

Although numerous studies have investigated the detailed distribution of
molecular gas in individual mergers (e.g. Yun \& Hibbard 2001;
Davis et al. 2013; Fernandez et al. 2014), a statistical study that
combines HI and H$_2$, with a well defined control sample is 
required in order to quantify general trends within mergers.
One possible reason for the lack of HI depletion in our merger sample,
despite elevated rates of star formation, could be that the stars
form from the extant molecular gas reservoir.  
However, the previous studies that have attempted to compare
mergers and isolated galaxies have found that the molecular fraction
is actually enhanced in mergers (Braine \& Combes 1993; Bettoni et al. 2001;
Stark et al. 2013).  
Further hints on the impact of the molecular gas reservoir in mergers
may be gleaned from observations of local
UV luminous galaxies, or `Lyman break analogs' (LBAs) selected to have 
properties similar to normal star-forming galaxies at high redshift 
(Heckman et al. 2005).  Interestingly, the
LBAs have many properties typical of local mergers, including high
SFRs, low metallicities and large IR luminosities (Hoopes et al. 2007; 
Overzier et al. 2011).  In a recent study of 6 LBAs, whose morphologies 
and emission
line kinematics again indicate interactions, Goncalves et al. (2014) 
found high CO fractions relative to `normal' low redshift galaxies
(Saintonge et al. 2011). Whilst some enhancement in the CO fraction 
may be expected from
the simple addition of reservoirs, as we found for the HI gas fractions,
it is also expected that the strong gas flows in mergers, that ultimately
drive star formation,  will also increase the CO fraction (Braine \& 
Combes 1993).  Conversely, Sargent et al. (2014) conclude that
starburst galaxies (which are not necessarily mergers) have CO fractions
that are typically 2--3 times lower than normal galaxies.

The measurement of normal HI masses, but elevated molecular fractions
in a small sample of mergers led Braine \& Combes (1993) to propose
that, contrary to depleting the gas supply, interactions might be 
responsible for the cooling of ionized
halo gas, which leads to a central build-up of H$_2$.  Such a mechanism
seems plausible theoretically, through the interaction of gas reservoirs  
(Marinacci et al. 2010; Moster et al. 2011; Tonnesen \& Cen 2012;  
Marasco et al. 2012).   Obtaining CO measurements for the post-mergers in our
sample, and studying simultaneously the changes in the atomic and
molecular gas components will therefore be an important complement to
the study presented here.

\section*{Acknowledgments} 

SLE and DRP acknowledge the receipt of an NSERC Discovery grant 
and DF and JLR acknowledge NSF grant AST-000167932 and a George Mason University 
Presidential Fellowship for support of this work. 
SLE is grateful to Trevor Mendel for discussions concerning stellar mass,
to the astronomers at CEA-Saclay for hospitality during the final
stages of this work and also to Paola Di Matteo, Gary Mamon, Florent
Renaud, Pierre-Alain Duc, Stephanie Juneau and Frederic Bournaud for interesting 
merger-related discussions during her visit.  We have also benefitted
from discussions with Romeel Dave, Mark Sargent and Sheila Kannappan.

\end{document}